\documentclass[12pt,a4paper]{article}
\usepackage{amsmath,amssymb,mathrsfs,framed,esint,slashed,upgreek,bbm}
\usepackage[colorlinks]{hyperref}
\usepackage[dvipsnames]{xcolor}
\usepackage[all]{xy}

\newlength{\bibitemsep}
\newlength{\bibparskip}
\let\oldthebibliography\thebibliography
\renewcommand\thebibliography[1]{%
  \oldthebibliography{#1}%
  \setlength{\parskip}{\bibitemsep}%
  \setlength{\itemsep}{\bibparskip}%
}

\DeclareMathAlphabet{\mathpzc}{OT1}{pzc}{m}{it}

\DeclareMathAlphabet{\mathbbmsl}{U}{bbm}{m}{sl}

\DeclareMathAlphabet{\mathbbmslbf}{U}{bbm}{bx}{sl}

\newcommand{\curl}{{\operatorname{curl}}}

\newcommand{\rem}[1]{}

\newcommand{\de}{{\rm d}}

\newcommand{\bz}{{\mathbf{z}}}
\newcommand{\bq}{{\boldsymbol{q}}}

\newcommand{\bv}{{\boldsymbol{v}}}
\newcommand{\bp}{{\boldsymbol{p}}}
\newcommand{\bk}{{\mathbf{k}}}

\newcommand{\bM}{{\mathbf{M}}}
\newcommand{\MCO}{{\bhJ_{\!\!\mathcal{M}}}}
\newcommand{\bX}{{\mathbf{X}}}

\newcommand{\bx}{{\boldsymbol{x}}}
\newcommand{\bs}{{\boldsymbol{s}}}

\newcommand{\bhX}{{\widehat{\bf X}}}

\newcommand{\bn}{{\boldsymbol{n}}}

\newcommand{\bsigma}{{\boldsymbol{\sigma}}}

\newcommand{\bmu}{{\boldsymbol{\mu}}}

\newcommand{\bw}{{\boldsymbol{w}}}

\newcommand{\bA}{{\mathbf{A}}}

\newcommand{\bE}{{\mathbf{E}}}
\newcommand{\bF}{{\mathbf{F}}}

\newcommand{\bhJ}{{\, \widehat{\!\boldsymbol{\cal J}\,}\!}}

\newcommand{\bu}{{\boldsymbol{u}}}

\newcommand{\beq}{\begin{equation}}
\newcommand{\eeq}{\end{equation}}
\newcommand{\ben}{\begin{eqnarray}}
\newcommand{\een}{\end{eqnarray}}
\newcommand{\bomega}{\boldsymbol{\omega}}

\renewcommand{\contentsname}{}

\numberwithin{equation}{section}

\begin{document}

\title{Madelung hydrodynamics of  spin-orbit coupling:\\action principles, currents, and correlations
}
\author{
	Cesare Tronci%$^{1,2}$
	\smallskip\\
	\small
	%$^1$
	\it School of Mathematics and Physics, University of Surrey, Guildford, United Kingdom\\
	\small
	%$^2$
	\it School of Physical and Mathematical Sciences, Nanyang Technological University, Singapore
}

\date{}

\maketitle

\begin{abstract}
We exploit the variational and Hamiltonian structures of quantum hydrodynamics with spin to unfold the correlation and torque mechanisms accompanying  spin-orbit coupling (SOC) in electronic motion. Using  Hamilton's action principle for the Pauli equation, we isolate SOC-induced quantum  forces that act on the orbital Madelung--Bohm trajectories and complement  the usual force terms known to appear in quantum hydrodynamics with spin. While the latter spin-hydrodynamic forces relate to the quantum geometric tensor (QGT),  SOC-induced orbital forces originate from a particular current operator that contributes prominently to the spin current. This distinction between   force terms  reveals two fundamentally different mechanisms generating quantum spin-orbit correlations. Leveraging the Hamiltonian structure of the  hydrodynamic system, we also elucidate spin transport features such as the correlation-induced quantum torques and the current shift in the spin Hall effect. This Hall shift leads to a  new definition of the transport spin current thereby addressing an open question in spintronics.  Finally, we illustrate the framework via the Madelung--Rashba equations for planar SOC configurations and propose a  particle closure for possible  numerical implementations.
\end{abstract}

\vspace{-.9cm}
{\footnotesize
\renewcommand{\contentsname}{} % Makes the title empty
\tableofcontents
}

\section{Introduction}
Despite the widespread interest in the role of spin-orbit coupling (SOC) and its planar Rashba version in spintronics and other fields \cite{BiNoVyChMa22,MaKoNiFrDu15,Szary}, a unified view of the various mechanisms underlying SOC dynamics remains elusive. Several SOC-induced phenomena are known, such as the appearance of the Hall effect \cite{DyPe71,EnRaHa07} and its associated anomalous velocity. {\color{black}However, many of these phenomena still lack a theoretical description and, for example, the underlying mechanisms of the spin Hall effect `remain mysterious and poorly understood' \cite{MaEtAl24}. Each SOC-induced process} is usually  studied on its own, without carrying over the entire self-consistent picture comprising both spin and orbital dynamics.  Indeed, unfolding all SOC-related features within one dynamical framework beyond the usual Schr\"odinger representation seems challenging. 
This paper takes a step in this direction by considering the hydrodynamic description of SOC within its underlying variational principle and Hamiltonian structure. This general setting allows the characterization of both the SOC-induced forces and the kinematic effects that affect the orbital Madelung-Bohm trajectories \cite{Bo82,Ma27} as a result of quantum spin-orbit correlations \cite{Nikolic1}. The latter are also responsible for quantum spin torques which are further elucidated by the {\color{black}proposed} treatment. In addition, the  hydrodynamic framework introduced here provides a new natural way to express the transport spin current in terms of the hydrodynamic velocity and the spin density, thereby addressing a currently open question in spin transport \cite{MaKiChIeSa23,Ra03,ShZhXiNi06,TaKaMi24}.

In the absence of SOC, the hydrodynamic treatment of particles with spin is due to Takabayasi \cite{Takabayasi1,Takabayasi2} and various revisitations  have appeared over the decades. Takabayasi's work also covered the relativistic treatment of quantum hydrodynamics \cite{Takabayasi3}, which provided the basis for current hydrodynamic approaches to the Dirac equation. {\color{black}For a recent account on this topic, see \cite{Fabbri}.
Here, we isolate} pure SOC effects in the absence of  magnetic fields and with no additional relativistic corrections. When  these become important, minimal coupling and the Darwin  term lead to additional features that further increase the level of complexity \cite{ChWu12,MoZaMaHe14}.

 The hydrodynamic picture of SOC presented here may be of interest in several contexts, from electron hydrodynamics in metals \cite{FrSc24,IdHaRaSc23} to spin-orbit-coupled Bose-Einstein condensates \cite{LiJiSp11}. Moreover, in spintronics the introduction of hydrodynamic variables such as density and velocity may lead to a convenient description of the charge current conversion into spin current. Plasmonics and, more generally, quantum plasmas provide additional contexts in which  a hydrodynamic picture of SOC may be beneficial \cite{AnTr18,ChWu12,Manfredi,MoZaMaHe14}, although in that case one needs to also include the self-consistent Maxwell dynamics of electromagnetic fields. 
 Since the hydrodynamics of  Dirac electrons has  appeared  in various semirelativistic regimes, our purpose  is not to simply present new sets of equations, but rather to  shed light on the  correlation forces and torques appearing in SOC dynamics by exploiting its underlying variational and Hamiltonian structures. As a result, a natural expression for the transport spin current is also provided. {\color{black}Unlike previous proposals, this new expression does not involve any simplifying approximations and retains consistency with the orthogonality condition between the spin torque and the spin density. {\color{black}In addition, the Fermi-Dirac equilibrium average of the new expression of the transport spin current exhibits a dependence on the electron concentration that was absent in previous proposals}.  
 Notice that this paper distinguishes between the \emph{spin current}, as ordinarily defined in the absence of SOC, and its necessary extension to SOC dynamics, that is the \emph{transport spin current}   \cite{Ra03}.  Also, we refer to spin current-densities simply as spin currents.
 
 }

\paragraph{Plan of the paper.}  The  Madelung transform is applied to  the Pauli equation with SOC in section \ref{sec:HydroSOC}, which introduces the full hydrodynamic treatment. The underlying variational principle is presented in section \ref{sec:Madelung}, while section \ref{sec:Mead} has a two-fold purpose.
On the one hand, it unfolds the spin current  decomposition between the semiclassical part and the  quantum part comprising quantum correlations. This quantum part is entirely determined by a current operator previously appeared in theoretical chemistry \cite{Me92} -- here referred to as the \emph{Mead current operator} -- and whose role {\color{black} in spin dynamics has recently been investigated in \cite{FarolfiEtAl}}. On the other hand, section \ref{sec:Mead} shows that the spin transport dynamics can be conveniently written in the frame moving with the orbital flow, in such a way that the variational principle can be expressed in terms of physically natural variables. This new variational  principle is formulated in section \ref{sec:varSOChydro}, which also presents the hydrodynamic equations and discusses the differences between SOC-induced forces and the other correlation force terms typically generated by the quantum geometric tensor. Section \ref{sec:discuss} presents a more thorough discussion with specific emphasis on the features appearing in the spin transport dynamics. Section \ref{sec:HamStr} exploits the Hamiltonian structure of SOC hydrodynamics to both highlight the role of the spin current in the orbital motion and unfold the current shift associated to the spin Hall effect. %The current shift presents a new term that  plays the role of a Hall current.
Section \ref{sec:circulation} continues the discussion by focusing on the role of the anomalous velocity introduced by SOC dynamics and presenting the evolution of the spin  geometric phase. The latter is generated by separate contributions involving both semiclassical  and quantum SOC terms as well as non-SOC quantum correlations. The quantum sources of SOC-induced geometric phase are entirely written in terms of the Mead current operator. Section \ref{sec:spinvec} moves on to presenting the spin-vector formalism and section \ref{sec:torques} discusses the distinction between semiclassical and quantum torques in the spin transport. Section \ref{sec:spincurr} shows how the Madelung picture  leads to a new definition of the transport spin current  whose Hall term differs from the expressions provided in  previous proposals, thereby making a contribution to an ongoing debate on the correct definition of {\color{black}transport} spin current.
Section \ref{sec:MadRas} specializes the treatment to consider the Madelung-Rashba equations for Rashba SOC in lower-dimensional configurations. While the SOC-induced forces on the  orbital trajectories exhibit a simpler form, the spin torques continue to reflect the complexity  characterizing SOC dynamics at all levels. {\color{black}In an attempt} to tackle this complexity, Section \ref{sec:bohmions}  formulates a  {\color{black}particle closure for possible numerical implementations} of SOC hydrodynamics. Extending previous work by the author \cite{FoHoTr19,HoRaTr21}, this  {\color{black}closure} is based on an ensemble of computational particles -- the \emph{bohmions} -- that sample the quantum hydrodynamic paths. Each particle carries a spin vector rendering the overall spin transport dynamics.

\section{Hydrodynamic equations with spin-orbit coupling\label{sec:HydroSOC}}

In this section, we apply the Madelung transform to the Pauli equation
\beq\label{Pauli}
i\hbar\frac{\partial \Psi}{\partial t}=
-\frac{\hbar^2}{2m}\Delta \Psi  -  i\frac{\hbar^2}{4m^2c^2}({\color{black}\bF\times\widehat{\bsigma}})\cdot\nabla\Psi+V\Psi,
\eeq
where $\bF=-\nabla V$ is the force generated by the external potential $V$ and $\widehat{\bsigma}=(\widehat{\sigma}_x,\widehat{\sigma}_y,\widehat{\sigma}_z)$ is the usual array of Pauli matrices. Notice that we avoid including electromagnetic fields in order to focus our discussion only on the specific effects of SOC. 

\subsection{Madelung form of the Pauli equation and its action principle\label{sec:Madelung}}
In the Pauli equation \eqref{Pauli}, the Madelung transform   may be applied to the orbital part of the Pauli spinor $\Psi$ by decomposing the latter as \cite{BBirula,Holland}
\beq\label{EF}
\Psi(\bx)=\psi(\bx,t)\chi(\bx,t)
\,,\qquad\text{with}\quad
\psi(\bx,t)=\sqrt{D(\bx,t)}e^{iS(\bx,t)/\hbar}
\quad\text{and}\quad
\|\chi(\bx,t)\|^2=1.
\eeq
Here, $\psi=\sqrt{D}e^{iS/\hbar}$ is a square-integrable Schr\"odinger wavefunction, while $\chi$ is a conditional spin state, that is a normalized complex vector depending parametrically on the spatial coordinates. In particular, the normalization $\|\chi(\bx,t)\|^2=\chi^\dagger(\bx,t)\chi(\bx,t)=1$ holds at each point $\bx$ in physical space and we emphasize that $\|\cdot\|$ denotes the standard norm on complex vectors.

Having introduced the orbital phase $S(\bx,t)$ and the probability density $D(\bx,t)\allowbreak=\|\Psi(\bx,t)\|^2\allowbreak=|\psi(\bx,t)|^2$, we proceed by seeking a set of hydrodynamic equations. While the standard procedure involves the direct substitution of \eqref{EF} in \eqref{Pauli}, here we wish to pursue an alternative methodology that exploits the action principle underlying the Pauli equation. In particular, we extend the treatment recently proposed in \cite{FoHoTr19} to account for SOC. Upon defining the incompressible \emph{SOC vector field}
\beq\label{Xdef}
\widehat{\bX}=\frac{\hbar}{4m^2c^2}\color{black}\bF\times \widehat{\bsigma},
\eeq
{\color{black}the action principle associated to \eqref{Pauli}} reads
\[
\delta{\int_{t_1}^{t_2}}\big(\langle\Psi,i\hbar\partial_t\Psi\rangle-h(\Psi)\big)\de t=0
,\qquad\text{where}\quad
h(\Psi)={\int}\bigg(\frac{\hbar^2}{2m}\|\nabla\Psi\|^2-{\hbar}\big\langle \Psi,i\widehat{\bX}\cdot\nabla\Psi\big\rangle+V\Psi\bigg)\de^3x
\]
is the total energy functional (Dirac Hamiltonian). {\color{black}Also}, $\langle\cdot,\cdot\rangle$ denotes the real part of the standard inner product $\langle\cdot|\cdot\rangle$ on complex vectors, {\color{black}so that}
\[
\langle\Psi_1,\Psi_2\rangle=\operatorname{Re}\langle\Psi_1|\Psi_2\rangle=\operatorname{Re}(\Psi_1^\dagger\Psi_2)
\] 
{\color{black}is a function of the spatial coordinates. Since the integrals are defined on the entire physical space, here we assume that the Pauli spinor $\Psi$ and its gradients have sufficiently fast decay for the integrals to be defined at all times.}
Now, inserting \eqref{EF} yields
\beq
\delta{\int_{t_1}^{t_2}}\big(D\langle\chi,i\hbar\partial_t\chi\rangle-D\partial_tS-h(D,S,\chi)\big)\de t=0,
\label{actprin}
\eeq
where the total energy is now expressed as
\beq
h(D,S,\chi)=
\int\!\left(\frac{D}{2m}{|\nabla S+{\bA}|^2}+ \frac{\hbar^2}{8m}\frac{(\nabla
  D)^2}{D}+D(\nabla S+\bA)\cdot\langle\widehat\bX\rangle+ D\,\epsilon(\chi,\nabla\chi)  \right)\de^3x
%  \\
%  =&
%\int\!\left(\frac{1}{2M}\frac{|\boldsymbol{\mu}+D\boldsymbol{\cal A}|^2}{D}+ \frac{\hbar^2}{8M}\frac{(\nabla
%  D)^2}{D}+ D\,\epsilon^*(\psi,\nabla\psi)  \right)\de^3r
   \label{EFHydroHamiltonian},
\eeq
{\color{black}and we have denoted}
\[
\langle \widehat{O}\rangle= \langle\chi{,} \widehat{O}\chi\rangle.
\]
Also, we have defined
\beq
\bA=\langle\chi,-i\hbar\nabla \chi\rangle
\qquad\text{and}\qquad
\epsilon(\chi,\nabla\chi)={V} -\langle\chi,\widehat{\bX}\cdot(i\hbar\nabla+\bA)\chi\rangle+ \frac{1}{2m}\left(\hbar^2\|\nabla\chi\|^2-A^2\right),
\label{defs}
\eeq
so that  $\bA$ is the Berry connection \cite{Berry} making the treatment 
invariant under the gauge freedom $(S,\chi)\mapsto (S+\varphi,\chi e^{-i \varphi/\hbar})$ introduced by the factorization \eqref{EF}; see \cite{AbediEtAl2012} for further discussions.
 Also, $\epsilon$ is a time-dependent effective potential whose gradient terms generate different types of quantum  correlations. We recall  the expression of the \emph{quantum geometric tensor} (QGT)    \cite{PrVa80}
 \beq\label{QGT}
 Q_{jk}=\hbar^2\langle\partial_j\chi|\partial_k\chi\rangle-A_jA_k,
 \eeq
so that the last term in the expression of $\epsilon$ reads $m^{-1}\operatorname{Tr}Q/2$. We observe the presence of the SOC (second)   term in $\epsilon$, which will play a predominant role in the remainder of this work.

\subsection{Mead current operator, spin current,  and  the orbital  frame\label{sec:Mead}}
At this point, one could continue by taking variations in \eqref{actprin} and expand all the various terms in the resulting  equations.  Taking arbitrary variations $\delta S$ in  \eqref{actprin}-\eqref{EFHydroHamiltonian} yields the following equation for the orbital density:
\beq\label{Deqn0}
\frac{\partial D}{\partial t}+m^{-1}{\operatorname{div}}\big(D(\nabla S+\bA+m\langle\widehat{\bX}\rangle)\big)=0.
\eeq
Instead, the variations $\delta D$ lead to the quantum Hamilton-Jacobi equation
\[
\frac{\partial S}{\partial t}+\frac1{2m}\big|\nabla S+\bA+m\langle\widehat\bX\rangle\big|^2=m|\langle\widehat\bX\rangle|^2-V_Q-\epsilon+\bigg\langle\chi,i\hbar\frac{\partial\chi}{\partial t}\bigg\rangle,
\]
where
\[
V_Q=-\frac{\hbar^2}{2m}\frac{\Delta\sqrt{D}}{\sqrt{D}}
\]
is the usual quantum potential responsible for wavepacket spreading and quantum interference \cite{Wyatt}.
In this section we will be particularly interested in the  spin evolution, which in the present context is found by isolating the  variations $\delta\chi$. After {\color{black}integrating by parts and a few rearrangements}, one has 
%\begin{multline*}
% i\hbar D(\partial_t{\chi} +  (\boldsymbol{u}+\widetilde{\bf X}-m^{-1}\bA)\cdot\nabla\chi )+\frac{i\hbar}{2}\operatorname{div}(D(\widetilde{\bf X}-m^{-1}\bA))\chi
% \\
% = mD(\bu-\langle\bhX\rangle-m^{-1}\bA)\cdot\widehat{\bf X}\chi
%% DV\chi+\frac{i\hbar}{2m}D\bA\cdot\nabla\chi+\frac{i\hbar}{2m}\operatorname{div}(D(i\hbar\nabla +\bA)\chi)
% +DV\chi-\frac{\hbar^2}{2m}\operatorname{div}(D\nabla\chi)
%\end{multline*}
%
\begin{multline}\label{chieqn0}
 i\hbar \frac{\partial \chi}{\partial t}+   \frac{i\hbar}m(\nabla S+m\bhX)\cdot\nabla\chi +
\frac{i\hbar}{2mD}{\operatorname{div}}\big(D(\nabla S+m\bhX)\big)\chi
 = (V+\nabla S\cdot\widehat{\bf X})\chi
-\frac{\hbar^2}{2mD}\operatorname{div}(D\nabla\chi).
\end{multline}
We observe that the present status of the treatment is not really enlightening and, as we will see, more insight can be obtained by  a close inspection of the general evolution laws for the various {\color{black}hydrodynamic quantities} {\color{black}in the Madelung picture}. 
To achieve this insight, we start by rewriting the function $\epsilon$ in terms of the conditional density matrix $\hat\rho=\chi\chi^\dagger$, not to be confused with the overall spin state $\hat\varrho={\int}D\hat\rho\,\de^3x$. Notice that this approach differs from the usual  spin vector formalism, which is dealt with in section \ref{sec:spinvec}. While these two pictures are equivalent, the density matrix is used in spintronics for different purposes \cite{BeVi08,MaNi13,NiDoPePe19}. More generally, the present density matrix formulation leaves the SOC vector field $\bhX$ arbitrary thereby allowing to  treat other types of SOC in e.g. spin-1 condensates \cite{SuQuXuZhZh16}, quantum nanowires \cite{GoZu02}, and planar nanostructures with finite width \cite{DyKa86}. 

The introduction of  $\hat\rho=\chi\chi^\dagger$ in the energy $\epsilon$ can be performed by using the relation $\hbar^2\|\nabla\chi\|^2-A^2=\hbar^2\|\nabla\hat\rho\|^2/2$  as well as the following result:
\begin{align*}\nonumber
\color{black}\big\langle\chi , \widehat{\bX} \cdot (\bA+i\hbar \,\nabla \big)\chi\big\rangle
=&\,
%\langle\chi|i\hbar \widehat{\bX} \cdot  \nabla \chi  \rangle
%+
% \langle \nabla \chi|i\hbar\chi\rangle \cdot \langle \widehat{\bX}\rangle
% \\=&\,
% i\hbar{\operatorname{Tr}}\big( \chi\chi^\dagger \widehat{\bX} \cdot  ( \nabla \chi\chi^\dagger+\chi \nabla \chi^\dagger) \big) 
%\\=&\,
%i\hbar\operatorname{Tr} ( \hat\rho \widehat{\bX} \cdot  \nabla \hat\rho  )
%\\=&
-\frac{i\hbar}2\operatorname{Tr} (  \widehat{\bX} \cdot  [\hat\rho,\nabla \hat\rho] )
\label{computations2}
,
\end{align*}
{\color{black}where the commutator arises by suitably projecting on the Hermitian part}.
Then, the time-dependent effective potential in \eqref{defs} becomes
\beq\label{neweps}
\epsilon(\chi,\nabla\chi)={V} 
+
\frac{\hbar}2{\operatorname{Tr}}(i\widehat{\bX}\cdot [\hat\rho,\nabla\hat\rho]) +\frac{\hbar^2}{4m}\|\nabla\hat\rho\|^2=\varepsilon(\hat\rho,\nabla\hat\rho).
\eeq
We observe that the energy density $D\varepsilon(\hat\rho,\nabla\hat\rho)$ carries the quantity 
 \beq\label{MeadPot}
{i\hbar}[\hat\rho,\nabla\hat\rho]
 \eeq
 in the  SOC energy. The latter accompanies the usual   term $m^{-1}\hbar^2D\|\nabla\hat\rho\|^2/4$ containing the QGT {\color{black}in \eqref{QGT}}, whose real part is now rewritten  as 
 \[
 \operatorname{Re}Q_{j\ell}=\frac12\operatorname{Tr}(\partial_j\hat\rho\,\partial_\ell\hat\rho).
 \] 
 We see that, while both the second and the third terms in \eqref{neweps} generate correlations via spatial inhomogeneities, the  third term carries two gradients and the SOC  term carries only one.
Notice that completing the square in \eqref{neweps} yields $\hbar^2\|\nabla\hat\rho\|^2+2m\hbar\operatorname{Tr}(i\nabla\hat\rho\cdot [\bhX,\hat\rho])=\|\hbar\nabla\hat\rho-im[\rho,\bhX]\|^2+m^2({\color{black}\langle|\bhX|^2\rangle-|\langle\bhX\rangle|^2})$, although here we prefer to retain the current form. 
The quantity \eqref{MeadPot} enjoys some remarkable geometric properties that first appeared {\color{black}in} Mead's work on molecular geometric phases \cite{Me92} and were recently used by the author to model the interaction dynamics of quantum and classical systems \cite{GBTr21}. The appearance of such a geometric quantity in the present  treatment confers SOC a geometric setting that is still to be  explored.

The quantity \eqref{MeadPot}  has a  physical interpretation in terms of quantum spin-orbit correlations. To see this, recall the expression $m^{-1}\hbar{\operatorname{Re}}(\Psi^\dagger\hat\bp\widehat{\bsigma}\Psi)/2$ of the spin current along with the usual momentum operator $\hat\bp=-i\hbar\nabla$  and let us use \eqref{EF} to write 
\beq\label{SpinCurr}
{\operatorname{Re}}(\Psi^\dagger\hat\bp\widehat{\bsigma}\Psi)=D(\nabla S+\bA)\langle\widehat{\bsigma}\rangle+m{\operatorname{Tr}}\big(\bhJ_{\!\!\mathcal{M}}\widehat{\bsigma}\big),
\qquad\text{where}\qquad
\bhJ_{\!\!\mathcal{M}}=\frac{i\hbar}{2m}D[\hat\rho,\nabla\hat\rho]
\eeq
will be referred to as the \emph{Mead current operator} (MCO). 
%We notice that the skew-symmetric part of $\MCO$ appears explicitly in the expression \eqref{EFHydroHamiltonian} of the total energy.
Thus, the spin current is naturally decomposed into a semiclassical mean-field  factorization and a purely quantum term containing the MCO, which itself is a multiple of the quantity  \eqref{MeadPot}. 
{\color{black}Notice that the spin current in \eqref{SpinCurr} corresponds to the ordinary definition normally adopted in the absence of SOC. In SOC dynamics, {\color{black}instead,} the transport spin current \cite{Ra03} involves an additional term which is the subject of an ongoing debate and will be the focus of later sections.}
%This decomposition unfolds the relation between the electric current, associated to the first term, and the overall spin current retaining the MCO contribution. 
%We emphasize that \eqref{SpinCurr} leads to an expression of the spin current that differs from that usually adopted in spintronics, which corresponds to the first semiclassical term and remains a subject of debate \cite{MaKiChIeSa23,Ra03,ShZhXiNi06,TaKaMi24}.
The decomposition in \eqref{SpinCurr} also reflects directly in the expression of the spin-orbit correlation, which is expressed {\color{black}here} as
\begin{multline*}
{\int}\Psi^\dagger\hat\bp\widehat{\bsigma}\Psi\,\de^3x-{\int}\Psi^\dagger\hat\bp\Psi\,\de^3x{\int}\Psi^\dagger\widehat{\bsigma}\Psi\,\de^3x
=
{\int} D(\nabla S+\bA)\langle\widehat{\bsigma}\rangle\de^3x
\\
-
{\int} D(\nabla S+\bA)\,\de^3x
{\int} D\langle\widehat{\bsigma}\rangle\,\de^3x
+m{\operatorname{Tr}}{\int} \MCO\widehat{\bsigma}\,\de^3x.
%+
%{\int} D{\operatorname{Tr}}\Big(\frac{i\hbar}2[\hat\rho,\nabla\hat\rho]\widehat{\bsigma}\Big)\de^3x.
\end{multline*}
While the first two terms on the right hand side comprise semiclassical correlations, the last term 
corresponds to quantum spin-orbit correlations  that are directly produced by the MCO, whose role {\color{black}in spin dynamics without SOC} has recently been investigated in \cite{FarolfiEtAl}.
 
The fact that $\bhJ_{\!\!\mathcal{M}}$ allows to express $\epsilon$   in terms of the spin density matrix  as in \eqref{neweps} unfolds another important property of the dynamics, which will be used directly in the next section. Indeed, upon defining the functional $\mathcal{F}={\int}D\epsilon(\chi,\nabla\chi)\,\de^3x={\int}D\varepsilon(\hat\rho,\nabla\hat\rho)\,\de^3x$, we can use  the chain-rule relation $\delta\mathcal{F}/\delta\chi=2(\delta\mathcal{F}/\delta\hat\rho)\chi$ so that  arbitrary variations $\delta\chi$ in  \eqref{actprin}-\eqref{EFHydroHamiltonian} yield \eqref{chieqn0} in the more suggestive form
\beq\label{chieqn}
 \partial_t{\chi} +  \boldsymbol{u}\cdot\nabla\chi 
 =
 \hat\xi\chi\,,
 \qquad\text{with}\qquad
 i\hbar\hat\xi= m\big(\bu-\langle\widehat{\bX}\rangle\big)\cdot\widehat{\bX}+ \frac{1}{D}\frac{\delta \cal F}{\delta \hat\rho},
\eeq
 where  we have defined the Madelung velocity
\beq\label{MadVel}
\bu=\frac1{m}(\nabla S+\bA)+\langle\widehat{\bX}\rangle.
\eeq
Without expanding further in \eqref{chieqn}, we observe that $\hat\xi$   identifies a space-dependent skew-Hermitian operator  on complex vectors. Thus, we can write the general evolution law for $\chi$ as
\beq
\label{chievol}
\chi(\bx,t)=U(\bx_0,t)\chi_0(\bx_0)|_{\bx_0=\boldsymbol\eta^{-1}(\bx,t)},
\eeq
where {\color{black}$\boldsymbol\eta(\bx,t)$ is the invertible Lagrangian path   generated by the Madelung velocity $\bu(\bx,t)$ via $\dot{\boldsymbol{\eta}}=\bu\circ\boldsymbol{\eta}$, while $U(\bx,t)$ is a space-dependent unitary propagator such that $\hat\xi=\dot{U} U^{-1}\circ\boldsymbol{\eta}^{-1}$ (here, $\circ$ denotes  composition of functions)}.
Equation  \eqref{chievol} means that, as  happens also in the absence of SOC \cite{FoHoTr19},  the  spin evolution naturally occurs in the frame moving with the orbital flow. This realization will allow us to write the full set of hydrodynamic equations in a convenient form, as presented in the next section.

A similar {\color{black}setting also applies to} the  orbital probability density obeying the continuity equation \eqref{Deqn0}, that is
\beq\label{Deqn1}
\partial_t D+\operatorname{div}(D\bu)=0.
\eeq
This means that  $D$ is  transported along the orbital flow, so that its evolution is given by the usual \emph{Lagrange-to-Euler  map}:
\beq\label{LtE}
D(\bx,t)={\int} D_0(\bx_0)\delta(\bx-\boldsymbol{\eta}(\bx_0,t))\,\de^3x_0.
\eeq
At this point, we will make use of the relations above in the action principle \eqref{actprin}-\eqref{EFHydroHamiltonian} in such a way to obtain {\color{black}a} complete set of hydrodynamic equations that is {\color{black}manifestly} independent of the gauge $\varphi$ associated to the symmetry $(S,\chi)\mapsto (S+\varphi,\chi e^{-i \varphi/\hbar})$ introduced by \eqref{EF}.

\subsection{Quantum hydrodynamics with spin-orbit correlations\label{sec:varSOChydro}}
We proceed to replace \eqref{Deqn1} and the first in \eqref{chieqn} in the action principle \eqref{actprin}-\eqref{EFHydroHamiltonian}.
After integrating by parts with respect to time the second term  in \eqref{actprin}, using  \eqref{chieqn}, \eqref{Deqn1}, and \eqref{MadVel} therein  yields
\beq
\delta{\int_{t_1}^{t_2}}\ell(\bu,D,\hat\xi,\hat\rho)\de t=0
,\quad\ \text{where}\quad\ 
\ell= {\int} D\bigg(\frac{m}2|\bu-\langle\widehat{\bX}\rangle|^2-\frac{\hbar^2}{8m}\frac{|\nabla
   D|^2}D+\hbar\langle i\hat\xi\rangle- \varepsilon(\hat\rho,\nabla\hat\rho)\bigg)\de^3x
\label{actprinEP}.
\eeq
Here, the Lagrangian $\ell$ comprises both orbital hydrodynamic terms containing the Madelung velocity $\bu$ and   quantum spin terms containing the density matrix $\hat\rho$. The latter arise from the $\chi$-terms in \eqref{actprin}; we refer to \cite{BLTr15,FoHoTr19} for similar treatments in the absence of SOC. Importantly,  $\ell$ is a Lagrangian functional of Euler-Poincar\'e type \cite{HoMaRa98,HoScSt09}, so that the variations $\delta D$, $\delta\bu$,  $\delta\hat\xi$, and $\delta\hat\rho$, must be calculated from \eqref{LtE} as well as the relations
\beq\label{varssss}
\bu(\bx,t)=(\dot{\boldsymbol{\eta}}\circ{\boldsymbol{\eta}}^{-1})(\bx,t)%=\dot{\boldsymbol{\eta}}({\boldsymbol{\eta}}^{-1}(\bx,t),t)
,\qquad\ 
\hat\xi(\bx,t)=(\dot{U} U^{-1}\circ\boldsymbol{\eta}^{-1})(\bx,t)%=\dot{U}(\bx_0,t) U(\bx_0,t)^{-1}|_{{\boldsymbol{\eta}}^{-1}(\bx,t)}
,\qquad\ 
\hat\rho(\bx,t)=(U\hat\rho_0U^{-1}\circ\boldsymbol{\eta}^{-1})(\bx,t)%=U(\bx_0,t)\hat\rho_0(\bx_0)U^{-1}(\bx_0,t)|_{{\boldsymbol{\eta}}^{-1}(\bx,t)}.
,
\eeq
where the last follows from \eqref{chievol} upon recalling $\hat\rho=\chi\chi^\dagger$. These relations are needed to express the dynamical variables in terms of the Lagrangian coordinates $(\boldsymbol{\eta}, U)$, whose variations $(\delta\boldsymbol{\eta}, \delta U)$ are arbitrary and vanish at the endpoints. Using \eqref{LtE} and \eqref{varssss} leads to
\beq
\delta D=-\operatorname{div}(D\bw)
,\qquad\qquad\ 
\delta\bu=\partial_t\bw+\bu\cdot\nabla\bw-\bw\cdot\nabla\bu,
\label{vars2_5}
\eeq
and
\beq
\delta { \hat\varrho} = [\zeta, {  \hat\varrho}] -  \bw\cdot\nabla \hat\varrho
,\qquad 
\delta \hat{\xi} = \partial _t \zeta + [ \hat\zeta , \hat{\xi}  ] + \bu  \cdot\nabla  \zeta   -  \bw\cdot\nabla   \hat{\xi}, 
\label{vars3}
\eeq
where 
$
\bw=\delta{\boldsymbol{\eta}}\circ{\boldsymbol{\eta}}^{-1}
$ and $
\hat\zeta=\delta{U} U^{-1}\circ\boldsymbol{\eta}^{-1}$ 
are arbitrary quantities vanishing at the endpoints.

After noticing that the first in \eqref{chieqn} leads to $(\partial_t+\bu\cdot\nabla)\hat\rho=[\hat\xi,\hat\rho]$, taking variations in \eqref{actprinEP} 
and rearranging using both \eqref{Deqn1} and
\[
{\operatorname{Tr}}\bigg( \bigg(D\frac{\partial \varepsilon}{\partial\hat\rho}-\partial_j\Big(D\frac{\partial \varepsilon}{\partial\partial_j\hat\rho}\Big)\bigg)\nabla\hat\rho\bigg)-\nabla\varepsilon
=
-D{\operatorname{Tr}}\bigg( \nabla \widehat{\bf X}\cdot \frac{\partial \varepsilon}{\partial \widehat{\bf X}}\bigg)-\partial_j{\operatorname{Tr}}\bigg( D\frac{\partial \varepsilon}{\partial \partial_j\hat\rho}\nabla\hat\rho\bigg)
\] 
 yields the following equations of quantum hydrodynamics with spin-orbit coupling:
\begin{align}\nonumber
&D(\partial_t+\bu\cdot\nabla) (\bu-\langle\bhX\rangle )
=
-m^{-1}D\nabla (V+V_Q)
-{\operatorname{Tr}}(\nabla\widehat{\bf X}\cdot\bhJ
%\bigg(m\hat\rho(\bu-\langle\widehat{\bf X}\rangle)+\frac{i\hbar}2[\hat\rho,\nabla\hat\rho]\bigg)
)
\\ &\hspace{9.25cm} 
-\partial_j {\operatorname{Tr}}\bigg(  \widehat{X}_j\MCO+\frac{\hbar^2}{2m^2}D \partial_j\hat\rho \nabla\hat\rho \bigg)
,
%-\frac\hbar2\nabla{\operatorname{Tr}}\big(\widehat{\bX}\cdot[iD\hat\rho,\nabla\hat\rho]\big)
%+\frac\hbar2{\operatorname{Tr}}\big(\widehat{\bX}\times\operatorname{curl}[iD\hat\rho,\nabla\hat\rho]\big)
\label{MomEqn}
\\
&i\hbar \big(\partial_t\hat\rho+\bu\cdot\nabla\big)\hat\rho
 =
%m \big[\widehat{\bf X}\cdot,%\bhJ
 %mD\hat\rho(\bu-\langle\bhX\rangle)+\frac{i\hbar}2D[\hat\rho,\nabla\hat\rho]
 %\big]
% +\frac{\hbar}{2}{\operatorname{div}}\bigg(\frac{\hbar}{m}D[\hat\rho,\nabla\hat\rho]-i\{\bhX,\hat\rho\}\bigg)
 %
 %+\frac{i\hbar}{2}{\operatorname{div}}\bigg[\hat\rho,[D\bhX,\hat\rho]-\frac{i\hbar}{m}D\nabla\hat\rho\bigg]
% ,
 %
 \bigg[m(\bu-\langle\widehat{\bf X}\rangle)\cdot\widehat{\bf X}+{\color{black}\frac{i\hbar}{2}\big[\nabla\hat\rho,\widehat{\bX}\big]+\frac{i\hbar}{2D}{\operatorname{div}}\Big(D\big[\hat\rho,\widehat{\bX}\big]+\frac{i\hbar}{m}D\nabla\hat\rho\Big)}
 ,\hat\rho\bigg]
 ,
\label{rhoeqn}
\\
&\partial_tD+\operatorname{div}(D\bu)=0.
\label{D-eqn}
\end{align}
Here, the current operator
\beq\label{defJ}
%V_Q=-\frac{\hbar^2}{2m}\frac{\Delta\sqrt{D}}{\sqrt{D}}
%\qquad\text{ and }\qquad
\bhJ =D\hat\rho(\bu-\langle\widehat{\bf X}\rangle)+\MCO
\eeq
%The first is the usual quantum potential from Madelung hydrodynamics that is responsible for wavepacket spreading and quantum interference \cite{Wyatt}. 
includes the MCO 
so that the  spin current may be written  as  $\hbar{\operatorname{Tr}}(\bhJ\widehat{\bsigma})/2$ using \eqref{SpinCurr}. Notice that  the notation adopted in this paper is such that 
\[
[\widehat{\bf A},\widehat{\bf B}]=[\widehat{A}_j,\widehat{B}_j],
\] 
{\color{black}with summation over repeated indexes} unless otherwise specified.

The hydrodynamic system \eqref{MomEqn}-\eqref{D-eqn} presents several features that deserve a discussion. For example, it is clear from equation \eqref{D-eqn} that the Madelung velocity \eqref{MadVel}  identifies the hydrodynamic velocity transporting the orbital probability $D$ and sweeping the spin  evolution across physical space. Also,  the hydrodynamic momentum equation \eqref{MomEqn} presents several features.  First, the momentum shift appearing on the left-hand side  is typical of SOC physics and has a classical correspondent in Thomas' precession \cite{Fr26,Thomas}. In addition, equation \eqref{MomEqn} unfolds different forces beyond the standard potential terms. Let us first consider the SOC force, {\color{black}which comprises  two terms:}
\[
-m{\operatorname{Tr}}(\nabla\widehat{\bf X}\cdot\bhJ)=
-mD\langle\nabla\widehat{\bf X}\rangle\cdot(\bu-\langle\widehat{\bf X}\rangle)
-m{\operatorname{Tr}}(\nabla\widehat{\bf X}\cdot\MCO)
.
\]
The first of these two terms  identifies a semiclassical inertial force associated to the  momentum shift by the quantity $-\langle\widehat{\bX}\rangle$. See \cite{LiFl92} for an account of the semiclassical features of SOC in the case of central forces. The second term $-m{\operatorname{Tr}}(\nabla\widehat{\bf X}\cdot\MCO)$ contains gradients of $\hat\rho$ and, similarly to the forces in the second line of \eqref{MomEqn}, is an essentially quantum term with no classical analogue. However, there is a distinction between the terms that involve the SOC vector field $\bhX$ and those that do not. Indeed, the former  are specifically responsible for the generation of {\color{black}SOC-induced}    correlations, which are identified by the presence of the MCO  discussed in Section \ref{sec:Mead}. These spin-orbit correlations  are distinctively different from those generated by the QGT, whose corresponding term $-m^{-2}{\hbar^2}\partial_j {\operatorname{Tr}}(D\nabla\hat\rho \partial_j\hat\rho)/2$ also appears in the absence of SOC \cite{FoHoTr19}. For example, while the latter includes both  first- and second-order gradients,  the second-order gradients in the SOC terms are not {\color{black}multiplied} by other spatial derivatives. In addition,  spin-orbit  terms depend directly on the force $\bF=-\nabla V$, which then plays a crucial role in governing correlation dynamics. The orbital wavepacket delocalization mechanisms produced by  these different types of quantum correlations  remain so far unexplored and deserve future investigations. It may be interesting to observe that {\color{black}the different} correlation mechanisms are governed by different geometric objects, that is the QGT and MCO, {\color{black}with} a distinct geometric nature. The QGT, for example, comprises a K\"ahler structure whose metric identifies the \emph{quantum Fisher information matrix} from quantum information theory \cite{LaSo23}. However, while the QGT has been widely investigated, the MCO requires further study.

SOC correlation terms are also identifiable in the evolution equation \eqref{rhoeqn} for the  spin density matrix $\hat\rho$. Indeed, on the one hand, the first term in the commutator corresponds to the semiclassical action of the spin-orbit operator $\bhX \cdot\hat\bp$ expressed in the hydrodynamic orbital frame via the replacement $\hat\bp\to\hat\bp + \bA$. On the other hand, the remaining terms in $\bhX$ are responsible for the production of purely quantum spin-orbit torques that combine with the very last term, arising  from the QGT. {\color{black}We observe that this last commutator term $ (\hbar^2/m)D^{-1}[\hat\rho,{\operatorname{div}}( D\nabla\hat\rho)]/2$ in \eqref{rhoeqn} can be regarded simultaneously as a quantum torque term and as a current term via the relation $i\hbar m^{-1}[\hat\rho,{\operatorname{div}}( D\nabla\hat\rho)]/2=\operatorname{div}\bhJ_{\!\!\mathcal{M}}$. Instead, the same argument does not apply to the quantum SOC  terms 
${i\hbar}[[\nabla\hat\rho,\widehat{\bX}]+D^{-1}\operatorname{div}[\hat\rho,D\widehat{\bX}] ,\hat\rho]/2$ in \eqref{rhoeqn}. This point will be crucial in  the characterization of current operators, as}
%Equation \eqref{rhoeqn} also possesses extra features that will be 
discussed in later  sections.

\section{Discussion\label{sec:discuss}}

Having presented the equations of hydrodynamics with SOC and illustrated the general meaning of the various terms, we move on to show how their structural properties shed light on various other features, especially concerning the spin transport dynamics.

\subsection{Spin Hall effect and Hamiltonian structure\label{sec:HamStr}}
As we will see in this and the next section, further insight into SOC-related effects can be obtained by moving to the Hamiltonian representation. This is written by introducing the canonical hydrodynamic momentum
\beq\label{momdef}
\bM=\frac{\delta\ell}{\delta \bu}=mD(\bu-\langle\widehat{\bX}\rangle)=D(\nabla S+\bA)
\eeq
and the weighted density matrix
\[
\tilde\rho=D\hat\rho=D\chi\chi^\dagger,
\]
so that the current operator $\bhJ$ in \eqref{defJ} is expressed as
\[
\bhJ=\frac1{mD}\Big(\tilde\rho\bM+\frac{i\hbar}{2}[\tilde\rho,\nabla\tilde\rho]\Big).
\]
With these variables, the Euler-Poincar\'e variational principle \eqref{actprinEP} is replaced by
\beq
\delta{\int_{t_1}^{t_2}}\!\bigg({\int} \big(\bM\cdot\bu+\langle\tilde\rho,i\hbar\hat\xi\rangle\big)\de^3x-h\bigg)\de t=0,
\label{actprinLP1}
\eeq
where
\beq
h= {\int} \bigg(\frac{|\bM|^2}{2mD}+\bM\cdot\langle\widehat{\bX}\rangle+\frac{\hbar^2}{8m\color{black} D}|\nabla
   D|^2+ D{\color{black}\mathscr{E}}(\tilde\rho,\nabla\tilde\rho)\bigg)\de^3x
\label{actprinLP2}
\eeq
and
\beq\label{neweps2}
{\color{black}\mathscr{E}}(\tilde\rho,\nabla\tilde\rho)={V} +\frac{\hbar}{2D^2}\langle\widehat{\bX},i[\tilde\rho,\nabla\tilde\rho]\rangle+ \frac{\hbar^2}{4mD^2}\big(\|\nabla\tilde\rho\|^2-|\nabla D|^2\big).
\eeq
Here, $h=h(\bM,D,\tilde\rho)$
is the (Dirac) Hamiltonian functional expressing the total energy \eqref{EFHydroHamiltonian} in the new set of variables, while $\color{black}\mathscr{E}(\tilde\rho,\nabla\tilde\rho):=\varepsilon(\tilde\rho/D,\nabla(\tilde\rho/D))$. 

Using \eqref{vars2_5}-\eqref{vars3}, taking arbitrary variations $\delta \bM$ in \eqref{actprinLP1} and denoting $\langle\widehat{O}\rangle=\operatorname{Tr}(\tilde\rho\widehat{O})/D$ yields the momentum equation \eqref{MomEqn} in the form
%\begin{multline}\label{CanMomEqn}
%\partial_t\bM+{\operatorname{div}}(\bu\bM)=-D{\nabla}(V+V_Q)
%-m{\operatorname{Tr}}(\nabla\widehat{\bf X}\cdot\bhJ
%)
%-\frac{i\hbar}{2}\partial_j {\operatorname{Tr}}\bigg(\frac1D\nabla\tilde\rho \bigg( [\widehat{X}_j,{\tilde\rho}]-\frac{i\hbar}{m}\partial_j\tilde\rho\bigg) \bigg)
%\end{multline}
%while the analogue of \eqref{rhoeqn}  reads
%\beq\label{CanRhoEqn}
%i\hbar \partial_t\tilde\rho+i\hbar\operatorname{div}(\tilde\rho\bu)
% =
%m \big[\widehat{\bf X}\cdot,\bhJ
% \big]
% -\frac{i\hbar}{2}{\operatorname{div}}\bigg(\frac1D\bigg[\tilde\rho,[\tilde\rho,\bhX]+\frac{i\hbar}{m}\nabla\tilde\rho\bigg]\bigg)
%\eeq
%and \eqref{D-eqn} is left unchanged.
%Rearranging to emphasize the role of the current operator $\bhJ$ yields the suggestive forms
\begin{multline}\label{CanMomEqn2}
\partial_t\bM+\partial_j \bigg(\frac{M_j\bM}{mD}\bigg)=D{\nabla}(V_Q-V)
-m{\operatorname{Tr}}(\nabla\bhX\cdot\bhJ)
+\partial_j {\operatorname{Tr}}\bigg(  \frac{\hbar^2}{2m}\tilde\rho\nabla(D^{-1}\partial_j\tilde\rho)-m\widehat{X}_j\bhJ\bigg)
,
\end{multline} 
while \eqref{rhoeqn} becomes
\beq\label{CanRhoEqn2}
i\hbar \partial_t\tilde\rho+{i\hbar}{\operatorname{div}}\big(\bhJ+\tilde\rho\langle\bhX\rangle\big)=\frac1D
\bigg[\bM\cdot\widehat{\bf X}+{i\hbar}
%{\color{blue}\Big[\nabla\tilde\rho+\tilde\rho\nabla {\ln}\sqrt{D},\widehat{\bX}\Big]}
%\qquad
\big[
\nabla\tilde\rho,\widehat{\bX}\big]-\frac{i\hbar}{2D}\big[\tilde\rho,\operatorname{div}( D\widehat{\bX})\big]
,\tilde\rho\bigg]
% m\big[\widehat{\bf X},\cdot\bhJ
% \big]
%  -\frac{i\hbar}{2}{\operatorname{div}}\bigg(\frac1D\big[\tilde\rho,[\tilde\rho,\bhX]\big]\bigg)
.
\eeq
%where $\{\cdot,\cdot\}$ denotes the anticommutator and we have used $[\tilde\rho,[\bhX,\tilde\rho]]=2D\langle\bhX\rangle\tilde\rho-D\{\tilde\rho,\bhX\}$. 
{\color{black}We} observe the minus sign in the last term above as well as the sign flip in the quantum potential term in \eqref{CanMomEqn2}.
We also rewrite equation \eqref{D-eqn} as
\beq\label{CanDEqn2}
\partial_tD+{\operatorname{div}}\big(m^{-1}\bM+D\langle\bhX\rangle\big)=0.
\eeq
The momentum equation \eqref{CanMomEqn2} once again unfolds the role of the current operator $\bhJ$ in generating quantum spin-orbit correlations. In particular, we observe that, in the case of spatially constant forces, one has $\nabla\bhX=0$ and  the evolution of the overall momentum expectation ${\int}\bM\de^3x$ does not depend explicitly on spin-orbit correlations.
 In addition, the  quantum  equation \eqref{CanRhoEqn2} reveals the nature of spin density transport. Indeed, we observe that the current operator is shifted by a \emph{Hall current} term that is nonlinear in the density matrix.
% appearing in the form of a \emph{torque dipole density} $\tilde\rho\langle\bhX\rangle$.  
As discussed in  \cite{CuSiSiJuMaNi04,ShZhXiNi06}, a shift in the current operator $\bhJ$  is a distinctive feature of the spin Hall effect in  spin state transport with SOC \cite{EnRaHa07,Gorini}. As we will see later, the Hall current operator 
\beq\label{HallCurrOp}
\color{black}\bhJ_{\!{\rm H}}=\tilde\rho\langle\bhX\rangle
\eeq
arising from the present Madelung approach differs from that appearing in the spintronics literature. {\color{black}The definition of the Hall  current in SOC dynamics is the subject of} a long-standing debate \cite{MaKiChIeSa23,Ra03,ShZhXiNi06,TaKaMi24}. In addition, the spin torques   in the  commutator {\color{black}on the right-hand side} of equation \eqref{CanRhoEqn2}  comprise both the first semiclassical SOC  term and the remaining  terms associated to quantum spin-orbit correlations. While we will come back to discussing  the spin torques in later sections, here we remark that equation \eqref{CanRhoEqn2} can be rewritten in such a way to further emphasize the role of the current operator $\bhJ$. Indeed, upon using 
%$\operatorname{div}\bhX=0$ to write
$D^{-1}\big[[\nabla\tilde\rho,\widehat{\bX}]-D^{-1}[\tilde\rho,\operatorname{div}( D\widehat{\bX})]/2,\tilde\rho\big]
 =
\big[ [\nabla\hat\rho,D\widehat{\bX}]+[\hat\rho,\operatorname{div}( D\widehat{\bX})
%{\color{blue}-D{\operatorname{div}}\bhX}
]/2,\hat\rho\big]
 $, the product rule and the Jacobi identity lead to
\beq\label{CanRhoEqn3}
i\hbar \partial_t\tilde\rho+{i\hbar}{\operatorname{div}}\Big(\bhJ+\frac12\{\tilde\rho,\bhX\}\Big)=
 m\big[\widehat{\bf X},\bhJ
 \big]
 %{\color{blue}+\frac1{2D}\big[\tilde\rho,[\tilde\rho,{\operatorname{div}}\bhX]\big]}
 .
\eeq
Here,  $\{\cdot,\cdot\}$ denotes the anticommutator and 
we have used 
$D[\hat\rho,[\bhX,\hat\rho]]=2\langle\bhX\rangle\tilde\rho-\{\tilde\rho,\bhX\}$.
This form shows that the reduced density matrix $\hat\varrho={\int}\tilde\rho\,\de^3x$ of the spin state evolves according to $i\hbar\de\hat\varrho/\de t=m{\int}[\widehat{\bf X},\color{black}\bhJ]\,\de^3x$, thereby revealing the predominant role of the current operator $\bhJ$ in the total spin dynamics. {\color{black}Importantly, we} notice that the right-hand side in \eqref{CanRhoEqn3}  does not identify a spin torque since it cannot be entirely expressed as a commutator having $\tilde\rho$ as one of its arguments. This means that, despite its suggestive appearance,  the anticommutator $\{\tilde\rho,\bhX\}/2$ cannot be interpreted as a spin current term. {\color{black}This point is crucial to isolate the correct expression of the Hall current term $\bhJ_{\!\mathrm{H}}$, as discussed in Section \ref{sec:spincurr} using the spin vector formalism}.
Moving to the orbital continuity equation \eqref{CanDEqn2}, we observe the presence of a shift in the hydrodynamic transport term that is analogous to the spin  case.

Equations \eqref{CanMomEqn2}-\eqref{CanDEqn2} possess a Hamiltonian structure with the Hamiltonian functional $h$ in \eqref{actprinLP2} and the Poisson bracket
\begin{align}\nonumber
\{f,k\}(\bM,\tilde\rho,D)=&{\int}\bM\cdot\bigg(\frac{\delta\color{black} f}{\delta \bM}\cdot\nabla\frac{\delta \color{black}k}{\delta \bM}-\frac{\delta \color{black}k}{\delta \bM}\cdot\nabla\frac{\delta \color{black}f}{\delta \bM}\bigg)\de ^3 x
-
{\int}\bigg(\frac{\delta f}{\delta \bM}\cdot\nabla\frac{\delta k}{\delta D}-\frac{\delta k}{\delta \bM}\cdot\nabla\frac{\delta f}{\delta D}\bigg)\de ^3 x
\nonumber
\\
&
-{\int}\bigg\langle \tilde\rho,i\hbar^{-1}\left[{\color{black}\frac{\delta f}{\delta \tilde\rho},\frac{\delta k}{\delta \tilde\rho}}\right]+\frac{\delta f}{\delta \bM}\cdot\nabla\frac{\delta k}{\delta \tilde\rho}-\frac{\delta k}{\delta \bM}\cdot\nabla\frac{\delta f}{\delta  \tilde\rho}\bigg\rangle\,\de^3  x.
\label{SDP-LPB-EF}
\end{align}
{\color{black}Here, no confusion should arise with the anticommutator notation $\{\widehat{A},\widehat{B}\}=\widehat{A}\widehat{B}+\widehat{B}\widehat{A}$ used \color{black}previously}.
As reported in \cite{FoHoTr19}, the same Poisson bracket appears in the absence of SOC. This Poisson bracket is Lie-Poisson \cite{HoScSt09,MaRa98} on the dual of the semidirect-product Lie algebra $\big(\mathfrak{X}(\Bbb{R}^3)\allowbreak\,\circledS\,\allowbreak\mathscr{F}(\Bbb{R}^3,\mathfrak{u}(\Bbb{C}^2))\big)\allowbreak\,\circledS\,\mathscr{F}(\Bbb{R}^3)$, where $\mathfrak{X}(\Bbb{R}^3)$ denotes the space of vector fields on the physical space $\Bbb{R}^3$, while $\mathscr{F}(\Bbb{R}^3)$ is the space of functions on $\Bbb{R}^3$ and the space $\mathscr{F}(\Bbb{R}^3,\mathfrak{u}(\Bbb{C}^2))$ comprises functions from $\Bbb{R}^3$ to the Lie algebra of  skew-Hermitian $2\times 2$ matrices. Lie-Poisson structures of this type are typical of complex-fluid hydrodynamics, as shown in \cite{GBRa2009,Holm02}.

The bracket \eqref{SDP-LPB-EF} possesses the following family of Casimir invariants:
\beq
C={\int} D \Phi(D^{-1}\tilde\rho)\de^3x,
\label{Casimirs1}
\eeq
for  any  real-valued analytic function $\Phi$. In the present context, this is the only available  nontrivial family of invariants since  the usual  helicity  associated to the vorticity $m^{-1}\curl(\bM/D-\bA)$ vanishes identically in the absence of quantum vortices, since $\bM/D-\bA=\nabla S$.

\subsection{Anomalous velocity, circulation, and geometric phase\label{sec:circulation}}

The orbital momentum dynamics in the Hamiltonian system  comprising \eqref{CanMomEqn2}, \eqref{CanRhoEqn2}, and \eqref{CanDEqn2} reveals a well-known property of SOC, that is the presence of an anomalous velocity perpendicular to the force $\bF=-\nabla V$ acting on the orbital dynamics. This feature is made explicit by dividing \eqref{CanMomEqn2} by $mD$ and introducing the \emph{canonical velocity}
\[
\bv=\frac{\bM}{mD}=\bu-\langle\widehat{\bX}\rangle=\frac1m(\nabla S+\bA)
\]
so that,  restoring the variable $\hat\rho=\tilde\rho/D$ and writing $\bhJ=D\hat\rho\bv+\MCO$, one obtains
\begin{equation}\label{v-eqn}
(\partial_t+\bu\cdot\nabla)\bv=
-\frac1m{\nabla}(V-V_Q)
-\frac1D{\operatorname{Tr}}(\nabla\bhX\cdot\bhJ)
-\frac1D\partial_j {\operatorname{Tr}}\bigg( \widehat{X}_j\MCO+\frac{\hbar^2}{2m^2}D\partial_j\hat\rho\nabla\hat\rho  \bigg).
\end{equation}
We observe  the  appearance in \eqref{v-eqn} of the so-called \emph{anomalous velocity}  $\langle\bhX\rangle={\hbar}\langle\widehat{\bsigma}\rangle\times\bF/({4m^2c^2})$ in the hydrodynamic transport term on the left-hand side, which  depends explicitly on the spin state variable $\hat\rho$.
The appearance of this anomalous velocity is known as a typical feature of SOC \cite{EnRaHa07}. It may be worth emphasizing that, as it appears from the equations \eqref{MomEqn}, \eqref{rhoeqn}, and \eqref{D-eqn}, SOC
hydrodynamics presents only one transport velocity, which is the Madelung velocity $\bu=\bv+\langle\bhX\rangle$
in \eqref{MadVel}, advecting the orbital density and sweeping the spin state across physical space. The anomalous character of the velocity term $\langle\bhX\rangle$ becomes manifest in the continuity equation \eqref{CanDEqn2} for the orbital density $D$, which shows how the usual probability current $m^{-1}\bM=D\bv$ must be completed by the SOC term $D\langle\bhX\rangle$ to realize the hydrodynamic transport. A similar argument applies to the evolution of the spin state density \eqref{CanRhoEqn2}.

%Indeed, this step leads to
%\begin{multline}
%m(\partial_t+{\bv}\cdot\nabla)\bv=
%-\nabla (V+V_Q)
%%-MD\langle\hat\rho,\pounds_{\widehat{\bX}}{\bv}\rangle
%+m\langle{\widehat{\bX}}\rangle\times\boldsymbol{\cal B}
%-m\langle\nabla(\bv\cdot\bhX)\rangle
%\\
%-\frac{\hbar^2}{2mD}\partial_j{\operatorname{Tr}}\big( D\nabla\hat\rho\partial_j\hat\rho\big)
%-\frac\hbar{2D}\nabla{\operatorname{Tr}}\big(\widehat{\bX}\cdot[iD\hat\rho,\nabla\hat\rho]\big)
%+\frac\hbar{2D}{\operatorname{Tr}}\big(\widehat{\bX}\times\operatorname{curl}[iD\hat\rho,\nabla\hat\rho]\big)
%\end{multline}

Equation \eqref{v-eqn} is also relevant in  unfolding the circulation balance, which reads as
\beq\label{circ-eqn}
\frac{\de}{\de t}{\oint_{c(t)}}\bv\cdot\de\bx=-{\oint_{c(t)}}
\bigg(\big\langle\nabla(\bv\cdot\bhX)\big\rangle
+\frac{1}{D}
{\operatorname{Tr}}\big(\nabla\bhX\cdot\MCO\big)
+\frac{1}{D}
\partial_j {\operatorname{Tr}}\bigg(  \widehat{X}_j \MCO+\frac{\hbar^2}{2m^2}D\partial_j\hat\rho\nabla\hat\rho  \bigg)\bigg)\cdot\de\bx,
\eeq
for any loop $c(t)$ moving along the orbital flow with the Madelung velocity $\bu=\bv+\langle\widehat{\bX}\rangle$. Due to the identity $\operatorname{curl}\bA=m\operatorname{curl}{\bv}$, we observe that \eqref{circ-eqn} actually identifies the evolution of the geometric phase $\oint_{c(t)}\bA\cdot\de\bx=m\oint_{c(t)}\bv\cdot\de\bx$. We {\color{black}notice} that the first term in \eqref{circ-eqn} corresponds to an averaged SOC force, while the remainder  of the circulation is generated by quantum correlations associated to both the MCO and the QGT.
%In the case of a constant force, that is $\nabla\bhX=0$, this geometric phase is generated only by  quantum spin-orbit correlations and is not driven by any semiclassical mechanism.  
In the  absence of SOC, the circulation balance \eqref{circ-eqn} recovers the geometric phase dynamics already found in \cite{FoHoTr19,MaBu24}.

\section{Spin-vector representation and Rashba dynamics\label{sec:spinvec}}
While the preceding treatment was performed using quantum operators, the spin-vector formalism may unfold other properties such as the  torques arising from spin-orbit correlations. The spin vector $\bs=\bs(\bx,t)$ is introduced as usual via the relation $\hat\rho=\color{black}\boldsymbol{1}/2+\bs\cdot\widehat{\bsigma}/\hbar$, where $\bs(\bx,t)=\hbar\langle\widehat{\bsigma}\rangle/2$ so that $|\bs(\bx,t)|=\hbar/2$. 

\subsection{Spin-orbit hydrodynamics and torques\label{sec:torques}}

{\color{black}In the remainder of this paper, we consider electronic motion under the the force $\bF$  given by an external electric field  $\bE=-\nabla\phi$, that is
$
\bF=-e\bE%=e\nabla\phi
$,
so that the definition \eqref{Xdef} yields
\beq
\langle\bhX\rangle
=\frac{e\hbar}{4m^2c^2}\langle\widehat{\bsigma}\rangle\times\bE
=\frac{e}{2 m^2c^2}\bs\times\bE
%=-\boldsymbol{\cal F}\bs
%\,,\qquad\text{with}\qquad
%{\cal F}_{ja}=-\frac{1}{2 m^2c^2}\upvarepsilon_{jak}F_k
.
\eeq
}Also, the quantity \eqref{MeadPot} becomes
$
{i\hbar}[\hat\rho,\nabla\hat\rho]
=-
{2}{\hbar}^{-1} \widehat{\sigma}_a \upvarepsilon_{a cb } s_c\nabla s_b
=
{2}{\hbar}^{-1} \widehat{\sigma}_a
(\nabla\bs\times\bs)_a
$,
so that the first relation in \eqref{SpinCurr} leads to writing the tensor components of the spin current as
\beq\label{spincurrent}
{\color{black}
\mathbbmsl{J}_{k a}}=\frac\hbar{2m}{\operatorname{Re}}(\Psi^\dagger\hat{p}_k\widehat{\sigma}_a\Psi)=D\big(  v_k s_a+ m^{-1}
(\partial_k\bs\times\bs)_a\big).
\eeq
We observe that, in the spin-vector representation, the MCO $\MCO$ in \eqref{SpinCurr}  is replaced by the current $D\nabla \bs\times\bs/m$, where the quantity $\nabla \bs\times\bs$ has previously appeared in the theory of complex fluids \cite{GBRaTr13,Holm02} under the name of \emph{wryness tensor} \cite{Eringen}.  
%Also, we emphasize that the expression \eqref{spincurrent} of the spin current extends the one commonly adopted in spintronics \cite{MaKiChIeSa23,Ra03,ShZhXiNi06,TaKaMi24} precisely by the addition of the second term, which is crucial to realize quantum spin-orbit correlations.
A useful quantity is the spin current vector 
with components $\color{black}J_{\!\ell}=\upvarepsilon_{\ell ka}\mathbbmsl{J}_{ka}$, that is
\beq\label{spinCurr}
{\color{black}\boldsymbol{J}}=\frac\hbar{2m}{\operatorname{Re}}(\Psi^\dagger\hat{\bp}\times\widehat{\bsigma}\Psi)=\frac1mD\big(m\bv\times\bs+\bs\cdot\nabla \bs- \bs\operatorname{div}\bs\big).
\eeq
Then, the SOC energy is expressed as
\[
{\int}\Psi^\dagger\bhX\cdot\hat{\bp}\Psi\,\de^3\bx
=
\color{black}
\frac{e\hbar}{4m^2c^2}{\int}\bE\cdot{\operatorname{Re}}(\Psi^\dagger\hat{\bp}\times\widehat{\bsigma}\Psi)\,\de^3\bx
%=
%D\boldsymbol{\cal F}\cdot(m\bv\times\bs+\bs\cdot\nabla \bs- \bs\operatorname{div}\bs)
=e\boldsymbol{\cal E}\cdot\boldsymbol{J},
\]
where we have defined
\[{\color{black}
\boldsymbol{\cal E}=\frac{1}{2mc^2}\bE.
%{\,\color{blue}=-\frac{1}{2m^2c^2}\nabla V}.
}
\]

%Also, we compute
%\begin{align*}
%\frac{\hbar}2{\operatorname{Tr}}(\bhX[i\hat\rho,\nabla\hat\rho])
%%=&\,
%%\frac{1}2{\operatorname{Tr}}(\widehat{X}_j[i\hbar\hat\rho,\partial_j\hat\rho])
%%\\
%%=&
%%\frac{1}{4m^2c^2} (\partial_j \bs\times\bs )_a\upvarepsilon_{jbn}{\operatorname{Tr}}(\widehat{\sigma}_a\widehat{\sigma}_b) F_n
%%\\
%=%&
%\frac{1}{2m^2c^2} \upvarepsilon_{jan} F_n\upvarepsilon_{abc} s_c\partial_j s_b 
%=
%\frac{1}{2m^2c^2}  F_n\upvarepsilon_{jan}(\partial_j\bs\times\bs)_a
%=
%2\boldsymbol{\cal F}\cdot\check{\bgamma}
%=
%(\bs\cdot\nabla\bs-\bs\,{\operatorname{div}}\bs)\cdot\boldsymbol{\cal F},
%\end{align*}
%where we have defined
%\[
%\gamma_{aj}=-\upvarepsilon_{abc} s_b\partial_j s_c
%\,,\qquad\text{and}\qquad
%\check{\gamma}_{n}=\frac14\upvarepsilon_{nja}(\bgamma-\bgamma^T)_{aj}=(\check{\gamma}_\text{\tiny Skew})_n
%\,,\qquad\text{and}\qquad
%\boldsymbol{\cal F}=\frac{1}{2m^2c^2}\bF.
%\]
%We see that retaining the spin current explicitly in the treatment requires dealing with the tensor $\pmb{\Bbb{F}}$, which makse the treatment rather cumbersome. For example, we have
%$
%\langle\bhX\rangle
%=\pmb{\Bbb{F}}\bs
%$.

Using these identities, 
%as well as
%\[
%\nabla\bs\cdot\bigg( D\frac{\partial \varepsilon}{\partial\bs}-{\operatorname{div}}\Big(D\frac{\partial \varepsilon}{\partial\nabla\bs}\Big)\bigg)-\nabla\varepsilon
%=
%-D\nabla\boldsymbol{\cal F}\cdot\frac{\partial \varepsilon}{\partial \boldsymbol{\cal F}} -{\operatorname{div}}\bigg( D\frac{\partial \varepsilon}{\partial \nabla{\cal F}_a}\nabla{\cal F}_a\bigg)
%,
%\]
 several rearrangements take
the hydrodynamic system \eqref{MomEqn}-\eqref{D-eqn} into 
\begin{align}\nonumber
&mD(\partial_t+\bu\cdot\nabla) \bv
=%&\,
-D\nabla ({\color{black}V_Q-e\phi})
-{\color{black}e\nabla\boldsymbol{\cal E}\cdot \boldsymbol{J}}
%(m\bv\times\bs+\bs\cdot\nabla\bs-\bs\,{\operatorname{div}}\bs)
\\&\,\hspace{6.75cm}
-{\color{black}\frac1m\partial_j\Big(D\nabla \bs\cdot\partial_j\bs+eD\big((\nabla\bs\times\bs)\times\boldsymbol{\cal E}\big)_j\Big)},
\label{MomEqn-spin}
\\\nonumber
&(\partial_t+\bu\cdot\nabla)\bs=\color{black}\frac{e}m\Big(
m\boldsymbol{\cal E}\times\bv+
2\nabla \bs\cdot\boldsymbol{\cal E}-2\boldsymbol{\cal E}\operatorname{div}\bs
\\&\,\hspace{6.75cm}{\color{black}
+\frac{1}{D}\bs\times{\operatorname{curl}}( D\boldsymbol{\cal E} )
-\frac1{eD}\operatorname{div}(D\nabla\bs)
\Big)}\times\bs,
\label{rhoeqn-spin}
\\
&
\partial_tD+\operatorname{div}(D\bu)=0,
\label{D-eqn-spin}
\end{align}
where  $\color{black}((\nabla\bs\times\bs)\times\boldsymbol{\cal E})_j=s_j\nabla\bs\cdot\boldsymbol{\cal E}-(\bs\cdot\boldsymbol{\cal E})\nabla s_j$ and the Madelung velocity now reads
\[
\bu=\bv+\color{black}\frac{e}m\bs\times\boldsymbol{\cal E}.
\]
{\color{black}Equations \eqref{MomEqn-spin}-\eqref{D-eqn-spin} are consistent with those found in \cite{MoZaMaHe14} within a more general  context.
Introducing the spin density $\tilde\bs=D\bs$ and the momentum density $\bM=mD\bv$}, one can also write the analogues of \eqref{CanMomEqn2}-\eqref{CanDEqn2} as follows:
\begin{align}%\nonumber
&\partial_t\bM+\partial_j\bigg(\frac{M_j\bM}{mD}\bigg)=
{\color{black}eD\nabla \phi
-e\nabla\boldsymbol{\cal E}\cdot \boldsymbol{J}}
%(mD\bv\times\tilde\bs+\tilde\bs\cdot\nabla\tilde\bs-\tilde\bs\,{\operatorname{div}}\tilde\bs)
%\\&\,\hspace{7cm}
+\partial_j\Big(\frac1m\tilde\bs\cdot\nabla (D^{-1}\partial_j\tilde\bs)-{\color{black}e(\mathbbmslbf{J}\times\boldsymbol{\cal E})}_j\Big),
\label{CanMomeqn-spin}
\\\nonumber
&
\partial_t\tilde\bs+{\operatorname{div}}\Big(\color{black}\mathbbmslbf{J}+\frac{e}{mD}({\tilde\bs\times\boldsymbol{\cal E}})\tilde\bs\Big)=\frac{e}{mD}\big(\boldsymbol{\cal E}\times\bM
+
2\nabla(\tilde\bs\cdot\boldsymbol{\cal E})
\\&\,\hspace{7.5cm}\color{black}
-2{\operatorname{div}}(\tilde\bs\boldsymbol{\cal E})
%\color{blue}
-D^{-1}\tilde\bs\times{\operatorname{curl}}( D\boldsymbol{\cal E} )
\big)\times\tilde\bs
,
\label{Canrhoeqn-spin}
\\
&
\partial_t D+\color{black}m^{-1}{\operatorname{div}}\big(\bM+e{\tilde\bs\times\boldsymbol{\cal E}}\big)=0,
\label{CanD-eqn-spin}
\end{align}
where the notation is such that, for an arbitrary vector $\bn$, we have $\color{black}\bn\cdot(\mathbbmslbf{J}\times\boldsymbol{\cal E})_j=n_i\upvarepsilon_{ j\ell k}\mathbbmsl{J}_{i\ell}{\cal E}_k$, and thus  $\color{black}m(\mathbbmslbf{J}\times\boldsymbol{\cal E})_j=m\bv(\tilde\bs\times\boldsymbol{\cal E})_j+D^{-1}((\nabla\tilde\bs\times\tilde\bs)\times\boldsymbol{\cal E})_j$. Notice that the quantum potential has now been absorbed in the third term on the right hand side of \eqref{CanMomeqn-spin} and no longer appears explicitly; see the change of variables in \cite{FoTr}. The momentum dynamics in either \eqref{MomEqn-spin} or \eqref{CanMomeqn-spin} again allows  the distinction between the correlations produced by the QGT (second to last term) and those produced by SOC (terms involving $\color{black}\boldsymbol{\cal E}$). 
In addition, we observe again the effect of the anomalous velocity $\color{black}m^{-1}e\tilde\bs\times\boldsymbol{\cal E}/D$ in the density transport \eqref{CanD-eqn-spin}. More importantly, the spin transport equation  \eqref{Canrhoeqn-spin} allows for an explicit characterization of the spin-orbit torques excerpted on the spin vector, which are are clearly identified by the $\color{black}\boldsymbol{\cal E}$-terms on the right-hand side in either \eqref{rhoeqn-spin} or \eqref{Canrhoeqn-spin}. Also, we recognize that the first of these terms corresponds to the semiclassical torque action of  SOC in the orbital frame, while the remaining terms on the right-hand side identify torques generated by spin-orbit correlations. Notice the presence of gradient contributions to the torque from both the spin density and the orbital density. 
One may also {\color{black}rewrite \eqref{Canrhoeqn-spin} in a form analogous to \eqref{CanRhoEqn3}, that is}
\beq\label{CanRhoEqn3-analogue}
\partial_t\tilde{s}_\ell+\partial_k\Big(\color{black}\mathbbmsl{J}_{k\ell}+\frac{e\hbar^2}{4m}\upvarepsilon_{k \ell j}D{\cal E}_j\Big)= e(\mathbbmsl{J}_{\ell k}
{\cal E}_k
-
\mathbbmsl{J}_{k k}
{\cal E}_\ell
).
\eeq
As discussed in section \ref{sec:HamStr}, and contrary to the conclusions in \cite{BuNuMa04,ErScLo05}, the parenthesis on the left-hand side cannot be identified as a transport spin current, {\color{black}since} the spin density equation above is not structured to identify the right-hand side as a torque. {\color{black}Indeed, we find {\color{black}that this right-hand side is not orthogonal to the spin density, since} $s_\ell(\mathbbmsl{J}_{\ell k}
{\cal E}_k
-
\mathbbmsl{J}_{k k}
{\cal E}_\ell
)=-m^{-1}{\hbar^2}D\boldsymbol{\cal E}\cdot
\operatorname{curl}\bs/2\neq0$.} Rather,  the transport spin current is determined  in the present context by the left-hand side of \eqref{Canrhoeqn-spin} and deserves a separate discussion, which is provided in the next section.

{\color{black}
We conclude this section by emphasizing that the Hamiltonian structure given by \eqref{SDP-LPB-EF} and \eqref{actprinLP2}-\eqref{neweps2} consistently transfers to the spin vector formalism for two-level quantum systems. In this case, the total energy Hamiltonian \eqref{actprinLP2} reads 
\beq
h(\bM,D,\tilde\bs)= {\int} \bigg(\frac{|\bM|^2}{2mD}-{eD\phi} +\frac{e}{mD}\bM\cdot\tilde\bs\times\boldsymbol{\cal E} 
+\frac{e}{mD}
\tilde\bs\cdot(\nabla\tilde\bs\cdot\boldsymbol{\cal E}-\boldsymbol{\cal E}\operatorname{div}\tilde\bs)
+ \frac{\|\nabla\tilde\bs\|^2}{2mD}\bigg)\de^3x,
\label{actprinLP2spin}
\eeq
while the Poisson bracket  \eqref{SDP-LPB-EF} becomes 
\begin{align}\nonumber
\{f,k\}(\bM,D,\tilde\bs)=&{\int}\bM\cdot\bigg(\frac{\delta \color{black}f}{\delta \bM}\cdot\nabla\frac{\delta \color{black}k}{\delta \bM}-\frac{\delta \color{black}k}{\delta \bM}\cdot\nabla\frac{\delta \color{black}f}{\delta \bM}\bigg)\de ^3 x
-
{\int}\bigg(\frac{\delta f}{\delta \bM}\cdot\nabla\frac{\delta k}{\delta D}-\frac{\delta k}{\delta \bM}\cdot\nabla\frac{\delta f}{\delta D}\bigg)\de ^3 x
\nonumber
\\
&
+{\int} \tilde\bs\cdot\bigg(\frac{\delta f}{\delta \tilde\bs}\times\frac{\delta k}{\delta \tilde\bs}-\frac{\delta f}{\delta \bM}\cdot\nabla\frac{\delta k}{\delta \tilde\bs}+\frac{\delta k}{\delta \bM}\cdot\nabla\frac{\delta f}{\delta  \tilde\bs}\bigg)\,\de^3  x.
\label{SDP-LPB-EFspin}
\end{align}
Then, equations \eqref{CanMomeqn-spin}-\eqref{Canrhoeqn-spin} arise by evaluating the functional derivatives in  the noncanonical Hamilton's equations
\begin{align*}
&\frac{\partial \bM}{\partial t}+{\operatorname{div}}\bigg(\frac{\delta h}{\delta \bM}\bM\bigg)+\bigg(\nabla\frac{\delta h}{\delta \bM}\bigg)\cdot\bM=-D\nabla\frac{\delta h}{\delta D}-\nabla\frac{\delta h}{\delta \tilde\bs}\cdot\tilde\bs
\\
&\frac{\partial \tilde\bs}{\partial t}+{\operatorname{div}}\bigg(\frac{\delta h}{\delta \bM}\tilde\bs\bigg)=\frac{\delta h}{\delta \tilde\bs}\times\tilde\bs
\,,\qquad\quad
\frac{\partial D}{\partial t}+{\operatorname{div}}\bigg(D\frac{\delta h}{\delta \bM}\bigg)=0.
\end{align*}
For example, if $\mathbbm{1}$ denotes the identity matrix, we have
\begin{align*}
\frac{\delta h}{\delta \tilde\bs}=
&\,
\frac{e}{mD}\boldsymbol{\cal E}\times\bM
+\frac{e}{mD}\left(\nabla\tilde\bs\cdot\boldsymbol{\cal E}-\boldsymbol{\cal E}\operatorname{div}\tilde\bs\right)
-{\operatorname{div}}\bigg(\frac{\tilde\bs\boldsymbol{\cal E}-(\tilde\bs\cdot\boldsymbol{\cal E})\mathbbm{1}+m^{-1}\nabla\tilde\bs}{D}
\bigg)
\\=
&\,
\frac{e}{mD}\big(\boldsymbol{\cal E}\times\bM
-2{\operatorname{div}}(\tilde\bs\boldsymbol{\cal E}-(\tilde\bs\cdot\boldsymbol{\cal E})\mathbbm{1})
-D^{-1}\tilde\bs\times{\operatorname{curl}}( D\boldsymbol{\cal E} )\big)
-
\frac1m{\operatorname{div}}(D^{-1}\nabla\tilde\bs).
\end{align*}
We observe the role of the functional derivatives in the transport equation for the spin density $\tilde\bs$: the Madelung hydrodynamic velocity  $\bu={\delta h}/{\delta \bM}$ intervenes in  the spin transport term, while ${\delta h}/{\delta \tilde\bs}$ determines the angular velocity of the spin precession. In this way, current and torque terms are directly related to each term in the total energy Hamiltonian \eqref{actprinLP2spin}. As already noticed in section~\ref{sec:varSOChydro}, the last energy term in \eqref{actprinLP2spin} produces a torque $-{\operatorname{div}}(D^{-1}\nabla\tilde\bs)\times\tilde\bs/m$ that can equivalently be regarded as a current term via the relation ${\operatorname{div}}(D^{-1}\nabla\tilde\bs)\times\tilde\bs={\operatorname{div}}(D^{-1}{\nabla\tilde\bs\times\tilde\bs})$. The associated current term  corresponds to the MCO in the  spin-vector representation. However,  such a dual torque/current role is not shared with any of the other  terms above.

\subsection{Nonlinear transport spin current and torque consistency\label{sec:spincurr}}
 As it appears from the spin density equation \eqref{Canrhoeqn-spin}, the  spin current $ \mathbbmslbf{J}$ in \eqref{spincurrent} is  shifted by the \emph{Hall spin current}
 \beq\label{HallSC}
 \mathbbmslbf{J}_{\rm H}=\frac{e}{mD}(\tilde\bs\times\boldsymbol{\cal E})\tilde\bs,
 \eeq
 which here has the role of the Hall current operator \eqref{HallCurrOp}. Then,  the overall \emph{transport spin current} has the nonlinear expression
\beq\label{trspicurrent}
 \mathbbmslbf{J}^{tr}= \mathbbmslbf{J}+ \mathbbmslbf{J}_{\rm H}
=
  \frac{1}{eD} \boldsymbol{J}_{\!c\,}\tilde\bs+ \mathbbmslbf{J}_{\!\cal M},
\eeq
where $ \boldsymbol{J}_{\!c\,}=eD\bu=e(D\bv+{e}\tilde\bs\times\boldsymbol{\cal E}/m)$ is the electric charge current density and we have defined the \emph{Mead spin current} 
\beq\label{MeadSC}
\mathbbmslbf{J}_{\!\cal M}=\frac1{mD}\nabla \tilde\bs\times\tilde\bs.
\eeq
As shown in \cite{FarolfiEtAl} in the absence of SOC, the dynamical effect of $\mathbbmslbf{J}_{\!\cal M}$   is to amplify spatial gradients in the spin orientation, thereby generating fine, small-scale structures that severely delocalize the precessional spin phase and drive the coupled system into a non-linear turbulent regime.
With the definition \eqref{trspicurrent}, the spin density transport reads
\beq\label{SpinTransportEq}
\frac{\partial \tilde\bs}{\partial t}+\operatorname{div} \mathbbmslbf{J}^{tr}=\bomega\times\tilde\bs
,\qquad\text{where}\qquad
\bomega=\frac{e}{mD}\bigg(\boldsymbol{\cal E}\times\bM
+
2\nabla(\tilde\bs\cdot\boldsymbol{\cal E})
-2{\operatorname{div}}(\tilde\bs\boldsymbol{\cal E})
-\frac{\tilde\bs}D\times{\operatorname{curl}}( D\boldsymbol{\cal E} )\bigg).
\eeq
The  equation of motion in \eqref{SpinTransportEq} is crucial in isolating the various terms governing spin transport. In particular, the definition of transport spin current is the subject of an ongoing debate \cite{MaKiChIeSa23,ShZhXiNi06,Sonin,TaKaMi24,Samokhin} and  the expression \eqref{trspicurrent} seems to appear here for the first time. Before proposing a definition of spin current, however,  it is important to focus on the spin torque.

One fundamental feature that seems to have gone overlooked in the spintronics literature is the fact that the torque  ${\bf T}=\bomega\times\tilde{\bs}$ is  orthogonal to the spin itself. While this is generally not true in the presence of an environment or any source of dissipation, the unitary dynamics considered here cannot lead to any rescaling of the spin vector $\bs=\tilde\bs/D$. Therefore, by direct comparison with the corresponding equation \eqref{rhoeqn-spin} for $\bs$, we conclude that the the spin torque identifies a purely precessional motion with the angular frequency $\bomega=m^{-1}e(D^{-1}
{\boldsymbol{\cal E}\times\bM}+
2\nabla \bs\cdot\boldsymbol{\cal E}-2\boldsymbol{\cal E}\operatorname{div}\bs
+{D}^{-1}\bs\times{\operatorname{curl}}( D\boldsymbol{\cal E} )
)$. 
In dealing with  unitary dynamics, once the precessional torque has been identified, the transport spin current can be determined by isolating the divergence terms. The importance of the spin torque in relation to the definition of the spin current has been {\color{black}considered} in \cite{SunXie}, although  its precessional character, i.e.
\beq\label{orthog}
{\bf T}\cdot\tilde{\bs}=0,
\eeq
was not discussed nor used in the treatment therein. 

A separate discussion is required about the role of the Mead spin current \eqref{MeadSC}. As we have seen in the previous section, the relation $\operatorname{div}\mathbbmslbf{J}_{\!\cal M}=m^{-1}\operatorname{div}(D^{-1}\nabla \tilde\bs)\times\tilde{\bs}$ naturally leads to the equivalent interpretation of $\operatorname{div}\mathbbmslbf{J}_{\!\cal M}$ as a quantum torque \cite{Holland}. This apparent arbitrariness in the interpretation  is generally eliminated  by resorting to the argument that, in the absence of both SOC and magnetic fields, the spin density is conserved exactly so that $\partial_t\tilde{\bs}+\operatorname{div}\mathbbmslbf{J}=0$, and  $\mathbbmslbf{J}_{\!\cal M}$ is simply interpreted as a `quantum torque contribution to the spin current' \cite{FarolfiEtAl}. Notice, however, that none of the other terms in \eqref{Canrhoeqn-spin} possesses this type of arbitrariness for {\color{black}a generic} electric field $\bE={2mc^2}\boldsymbol{\cal E}$. Indeed, as shown by equation \eqref{CanRhoEqn3-analogue}, using the product rule for the divergence operator in \eqref{SpinTransportEq} yields  non-precessional torque terms violating the  orthogonality relation \eqref{orthog}.  Nevertheless, some authors insist that any torque and  spin current definitions contain a level of arbitrariness,   because one could rewrite a torque  in the form of a divergence \cite{ShZhXiNi06,Sonin}, e.g. by fixing  a set of  conditions to explicitly solve the equation $\operatorname{div}\mathcal{J}=\bomega\times\tilde\bs$ for $\mathcal{J}$. As a simpler example, the charge current $\boldsymbol{J}_{\!c\,}=eD\bu$ itself is only defined up to an additive solenoidal term; see \S115 in \cite{LaLi77}. Since this type of argument seems to apply just as well to all  continuum systems with and without orientational order,  we will not deal with this particular aspect.
Here, the transport spin current  $ \mathbbmslbf{J}^{tr}$ in \eqref{trspicurrent} was derived  exactly from Hamilton's principle for Madelung hydrodynamics with SOC and  ensures a purely precessional spin torque ${\bf T}=\bomega\times\tilde\bs$ given in \eqref{SpinTransportEq}, thereby {\color{black}retaining} physical consistency.
%While the latter was derived in the context of unitary dynamics, it is important to emphasize that any universal definition of transport spin current  must remain valid regardless of the presence of dissipation. 
In addition, we notice that, in the infinite medium, the transport spin current  \eqref{trspicurrent} satisfies $\int\operatorname{div}\mathbbmslbf{J}^{tr}\,\de^3x=0$, which is another necessary consistency relation.

We now compare $\mathbbmslbf{J}^{tr}$ to other definitions of transport spin current appearing in the literature.
%\subsection{\color{black}Comparison with previous definitions of transport spin current }
Evidently,   \eqref{trspicurrent} differs from the conventional expression used in  spintronics  \cite{NiZaSo06,Ra03}, which  reads $\mathscr{J}{=\hbar}\Psi^\dagger\{\hat\bv,\widehat\bsigma\}\Psi/4$. Here, $\Psi$ is the original Pauli spinor  and the velocity operator $\hat\bv=i\hbar^{-1}[\widehat{H},\hat\bx]$ is   determined by the Hamiltonian operator $\widehat{H}=m^{-1}|\hat\bp|^2/2+V(\hat\bx)+\bhX\cdot\hat\bp$ such that the Pauli  equation \eqref{Pauli} is written as $i\hbar\partial_t\Psi=\widehat{H}\Psi$ \cite{NiZaSo06}. Since $i\hbar^{-1}[\bhX\cdot\hat\bp,\hat\bx]=\bhX$, the factorization  \eqref{EF} leads to $\hbar\Psi^\dagger\{\hat\bv,\widehat\bsigma\}\Psi/4=\mathbbmslbf{J}+D\hbar\langle\{\bhX,\widehat\bsigma\}\rangle/4$ and the standard properties of the Pauli matrices yield \cite{Sonin}
\beq\label{convSC}
\mathscr{J}_{jk}=\frac\hbar4\Psi^\dagger\{\hat{v}_j,\widehat\sigma_k\}\Psi=\mathbbmsl{J}_{jk}+\frac{e\hbar^2}{4m}\upvarepsilon_{jk\ell} D\mathcal{E}_\ell.
\eeq
Here, the resulting Hall spin current $m^{-1}e\hbar^2\upvarepsilon_{jk\ell} D\mathcal{E}_\ell/4$  differs substantially from that in \eqref{HallSC}. In particular, we observe that the latter allows for parallel components to the electric field which are instead prevented by the last term above.
Importantly, the spin current expression  $\mathscr{J}$ coincides with the parenthesis on the left-hand side of the spin density equation \eqref{CanRhoEqn3-analogue}. As noticed in \cite{BuNuMa04,ErScLo05}, this equation  may be rewritten  as $\partial_t\tilde{s}_\ell+\partial_k\mathscr{J}_{k\ell}= e(\mathscr{J}_{\ell k}
{\cal E}_k
-
{\cal E}_\ell\mathscr{J}_{k k}
)$, whose right-hand side does not identify an orthogonal spin torque.
Therefore, we argue that the quantity $\mathscr{J}$  leads to an unsuitable  definition of the transport spin current.

A proposal to rectify the conventional  definition is found in \cite{ShZhXiNi06}, where  the following expression is introduced:
\beq\label{currprop}
\frac\hbar4\Psi^\dagger\big(\{\hat{v}_j,\widehat\sigma_k\}+\{\hat{x}_j,\widehat{\tau}_k\}\big)\Psi
=
%J_{jk}+\frac{\hbar^2}4\upvarepsilon_{jk\ell} D\mathcal{F}_\ell
\mathscr{J}_{jk}
+ex_j(\mathscr{J}_{k\ell}
{\cal E}_\ell
-
{\cal E}_k\mathscr{J}_{\ell\ell}).
\eeq
Here,  $\widehat{\boldsymbol\tau}=i\hbar^{-1}[\widehat{H},\widehat\bsigma]$ is the torque operator and the term $\hbar\Psi^\dagger\{\hat{\bx},\widehat{\boldsymbol\tau}\}\Psi/4$ is known as \emph{torque dipole density} \cite{CuSiSiJuMaNi04}. {\color{black}The operator $\{\hat{\bv},\widehat\bsigma\}+\{\hat{\bx},\widehat{\boldsymbol\tau}\}$ in \eqref{currprop} can be  conveniently written in the Heisenberg representation: if $\hat{\bx}_{\sf H}$ and $\widehat\bsigma_{\sf H}$ are Heisenberg position and Pauli matrix operators, then the Heisenberg operator corresponding to $\{\hat{\bv},\widehat\bsigma\}+\{\hat{\bx},\widehat{\boldsymbol\tau}\}$ reads $\de\{\hat\bx_{\sf H},\widehat\bsigma_{\sf H}\}/\de t$, which is to be contrasted with the operator $\{\de\hat\bx_{\sf H}/\de t,\widehat\bsigma_{\sf H}\}$, whose Schr\"odinger-picture correspondent $\{\hat{\bv},\widehat\bsigma\}$ appears in \eqref{convSC}}.
The equality in \eqref{currprop} is verified as follows. We compute $\widehat{\boldsymbol\tau}=i\hbar^{-1}[{\bhX\cdot\hat\bp},\widehat\bsigma]=-m^{-1}e\widehat\bsigma\times(\,\widehat{\!\boldsymbol{\cal E}\,}\!\times\hat\bp)$ and it is important to notice that $\Psi^\dagger\widehat{\boldsymbol\tau}\Psi$  is not orthogonal to the spin density $\tilde\bs=\Psi^\dagger\widehat{\boldsymbol\sigma}\Psi$, and thus cannot identify a spin torque. This fact should come as no surprise: in general, the expectation of a  precessional torque operator is not  a precessional torque itself. With this in mind, we evaluate $\{\hat{x}_j,\widehat{\boldsymbol\tau}\}=-m^{-1}e
\widehat\bsigma\times(\,\widehat{\!\boldsymbol{\cal E}\,}\!\times\{\hat{x}_j,\hat\bp\})
$ and a few algebraic steps lead to $\hbar\Psi^\dagger\{\hat{x}_j,\widehat{\tau}_k\}\Psi/4
=
ex_j(\mathbbmsl{J}_{k\ell}
{\cal E}_\ell
-
{\cal E}_k\mathbbmsl{J}_{\ell\ell}
)$, where we recognize the appearance of the right-hand side of \eqref{CanRhoEqn3-analogue}. Then, the role of the spin current expression \eqref{currprop} may be unfolded by writing the exact equation \eqref{CanRhoEqn3-analogue} as
\beq\label{CanRhoEqn3-analogue2}
\partial_t\tilde{s}_\ell+\partial_k\big(\mathscr{J}_{k\ell}+ex_j(\mathscr{J}_{k\ell}
{\cal E}_\ell
-
{\cal E}_k\mathscr{J}_{\ell\ell})\big)= 
4e(\mathscr{J}_{\ell k}
{\cal E}_k
-
\mathscr{J}_{k k}
{\cal E}_\ell
)
+
e\bx\cdot\nabla(\mathscr{J}_{\ell k}
{\cal E}_k
-
\mathscr{J}_{k k}
{\cal E}_\ell
)
.
\eeq
As in the case of equation \eqref{CanRhoEqn3-analogue}, the right-hand side above is not orthogonal to the spin vector in the general case and therefore, despite its widespread use, it appears that the expression \eqref{currprop} can hardly be identified as a transport spin current, at least in the context of unitary SOC dynamics. The validity of the  proposal \eqref{currprop}  has been discussed in \cite{BrNu09} and more recently in \cite{TaKaMi24}.

%\begin{framed}
%Take the conventional current with an electric field along x: 
%\[
%\mathscr{J}_{y3}=\mathbbmsl{J}_{y3}+\frac{e\hbar^2}{4m}\upvarepsilon_{y31} D\mathcal{E}_1
%=\frac{1}{eD} {J}_{\!cy\,}\tilde{s}_3+\frac1{mD}(\tilde{s}_2\partial_y \tilde{s}_1-\tilde{s}_1\partial_y \tilde{s}_2)+\frac{e\hbar^2}{4m} D\mathcal{E}_1
%\]
%We expand $(D_0+\delta D)^{-1}=D_0^{-2}(D_0-\delta D)$, so that
%\begin{align*}
%\mathscr{J}_{y3}
%=&\,
%\frac{1}{e D_0^2}(D_0\tilde{s}_{03}+D_0\delta\tilde{s}_{3}-\tilde{s}_{03}\delta D) {\delta J}_{\!cy\,}
%+\frac1{mD_0^2}(D_0\tilde{s}_{02}+D_0\delta\tilde{s}_{2}-\tilde{s}_{02}\delta D)\partial_y \delta\tilde{s}_1
%\\&\,
%-\frac1{mD_0^2}(D_0\tilde{s}_{01}+D_0\delta\tilde{s}_{1}-\tilde{s}_{01}\delta D)\partial_y \delta\tilde{s}_2+\frac{e\hbar^2}{4m} (D_0+\delta D)\mathcal{E}_1
%\\
%=&\,
%\frac{1}{e D_0}\tilde{s}_{03} {\delta J}_{\!cy\,}
%+\frac1{mD_0}\tilde{s}_{02}\partial_y \delta\tilde{s}_1
%-\frac1{mD_0}\tilde{s}_{01}\partial_y \delta\tilde{s}_2+\frac{e\hbar^2}{4m} (D_0+\delta D)\mathcal{E}_1
%\\
%=&\,
%\frac1{mD_0}\tilde{s}_{02}\partial_y \delta\tilde{s}_1
%-\frac1{mD_0}\tilde{s}_{01}\partial_y \delta\tilde{s}_2+\frac{e\hbar^2}{4m} (D_0+\delta D)\mathcal{E}_1
%\end{align*}
%\end{framed}

Based on the above arguments, we are led to conclude that Madelung hydrodynamics singles out \eqref{trspicurrent} as a natural candidate for the transport spin current, whose expression does not seem to be found elsewhere in the literature. Importantly, the nonlinear character of the Hall spin current  \eqref{HallSC} in the expression $\mathbbmslbf{J}^{tr}= \mathbbmslbf{J}+ \mathbbmslbf{J}_{\rm H}$ of the transport spin current prevents the use of standard linear-response approaches in the calculation of key quantities such as the spin Hall conductivity \cite{Sinova}, which is also subject to debate \cite{Gorini,RaScGoVi12}. 
A further validation of the new current definition  requires investigations in the context of spintronics applications that fall outside the purpose of this work. In the Appendix, we have followed  Sonin's work \cite{Sonin1} to present the Fermi-Dirac averages of the equilibrium current for the new definition. The main result is a linear dependence of the two-band  average on the Fermi energy. This reflects a dependence on the electron concentration that seems to be missing in the corresponding result for the conventional definition, as recently argued  in \cite{Samokhin}.

}

\subsection{Madelung-Rashba equations\label{sec:MadRas}}

%Despite the simplification offered by the passage to the spin-vector formalism, these equations  still reflect the  challenging intricacies of SOC dynamics. Some simplifications are obtained in the case where the force is spatially constant, which is the case of 2D Rashba dynamics. 

Given the intricate nature of spin-orbit correlation terms in both the spin and hydrodynamic momentum evolution, here we  consider the simplifications offered by {\color{black}lower-dimensional Rashba dynamics. The standard planar case} is given upon replacing the SOC vector field \eqref{Xdef}  by 
\[
\bhX_{R}=\frac{\alpha_R}\hbar\widehat{\bsigma}\times\bz
.
\]
Here, $\bz$ is the unit vertical and $\alpha_R$ is the Rashba parameter. First proposed in the context of two-dimensional electron gases \cite{ByRa84}, this type of SOC  allows  the orbital dynamics to lie entirely in the horizontal plane thereby simplifying the treatment. In addition, the Rashba operator $\bhX_R\cdot\hat\bp$ is generated by  a spatially constant vertical electric field, whose intensity can be adjusted to vary the coupling magnitude given by $\alpha_R$ \cite{Rashba06}. For these reasons, Rashba dynamics offers a convenient framework for exploring SOC properties and is widely used in spintronics as well as other contexts in quantum technology \cite{BiNoVyChMa22,MaKoNiFrDu15}. {\color{black}See also \cite{Aires} for a full relativistic treatment of Dirac-Rashba dynamics and its  accompanying spin Hall effect.}
%In addition, as customary in the study of Rashba systems, we will also discard the potential $V$ in the original quantum dynamics \eqref{Pauli}.

The first simplification occurs at the level of the momentum equation \eqref{MomEqn}, which becomes
\beq\label{MomEqn-R}
mD(\partial_t+\bu\cdot\nabla) \bv
=%&\,
-D\nabla {\color{black}(V+V_Q)}
-\partial_j\Big(\frac1mD\nabla \bs\cdot\partial_j\bs+\upalpha D\big(s_j\nabla s_z- s_z\nabla s_j\big)\Big),
\eeq
where {\color{black}all differential operators are on the $xy$-plane, $V(x,y)$ is an external confining potential, and}
\[
\bu=\bv+\upalpha\bs\times\bz
\,,
\qquad\qquad
\upalpha=\frac{2\alpha_R}{\hbar^2}
.
\] 
{\color{black}Equation \eqref{MomEqn-R} above reveals the complete absence of semiclassical SOC-induced forces in its right-hand side. Also, 
%That is, orbital Rashba dynamics is governed exclusively by quantum correlations that are generated by a combination of the QGT and  SOC.  
%A possible future study of SOC correlations could be performed by omitting the $\hbar^2$-terms in \eqref{MomEqn-R} (including $V_Q$) and \eqref{rhoeqn-R} in order to isolate exclusively the contributions generated by the SOC mechanism.
we} emphasize that the geometric phase dynamics resulting from \eqref{MomEqn-R} simplifies considerably with respect to the previous relation \eqref{circ-eqn}, which now becomes
\beq\label{circ-eqn-R}
\frac{\de}{\de t}{\oint_{c(t)}}\bv\cdot\de\bx=-{\oint_{c(t)}}\bigg(\upalpha\nabla\bv\cdot(\bs\times\bz)+\frac1{mD}
\partial_j\Big(\frac1mD\nabla \bs\cdot\partial_j\bs+\upalpha D\big(s_j\nabla s_z- s_z\nabla s_j\big)\Big)\bigg)\cdot\de\bx.
\eeq
%Once again, we recognize that the geometric phase is entirely produced by quantum correlations.
The remaining equations \eqref{rhoeqn-spin}-\eqref{D-eqn-spin} read
\begin{align}
&(\partial_t+\bu\cdot\nabla)\bs=\upalpha\Big(
m \bz\times\bv+2
\nabla s_z-2\bz\operatorname{div}\bs
-\bs\times(\bz\times\nabla {\ln} D)
-\frac1{\upalpha m D}\operatorname{div}(D\nabla\bs)
\Big)\times\bs,
\label{rhoeqn-spin-R}
\\
&
\partial_tD+\operatorname{div}(D\bu)=0,
\label{D-eqn-spin-R}
\end{align}
The hydrodynamic equations  \eqref{MomEqn-R} and \eqref{rhoeqn-spin-R}-\eqref{D-eqn-spin-R} comprise the \emph{Madelung-Rashba system}.

The  Hall current displacement and the spin-orbit torques become particularly evident in terms of the density variables, that is momentum $\bM=mD\bv$ and spin density  $\tilde\bs=D\bs$. The analogues of \eqref{CanMomEqn2}-\eqref{CanDEqn2} read as follows:
\begin{align}
&\partial_t\bM+\partial_j\bigg(\frac{M_j\bM}{mD}\bigg)=-D\nabla V+
%D\nabla V_Q+
\partial_j\Big(\frac1m\tilde\bs\cdot\nabla (D^{-1}\partial_j\tilde\bs)-m\upalpha(
{\color{black}\mathbbmslbf{J}}\times\bz)_j\Big),
\label{CanMomeqn-spinR}
\\
&
\partial_t\tilde\bs+{\operatorname{div}}\big({\color{black}\mathbbmslbf{J}}+\upalpha D^{-1}({\tilde\bs\times\bz})\tilde\bs\big)=\upalpha D^{-1}\big(\bz\times\bM
+
2\nabla\tilde s_z-2\bz{\operatorname{div}}\tilde\bs
+
\tilde{\bs}\times(\bz\times\nabla {\ln} D)
\big)\times\tilde\bs
,
\label{Canrhoeqn-spinR}
\\
&
\partial_t D+{\operatorname{div}}\big(m^{-1}\bM+\upalpha {\tilde\bs\times\bz}\big)=0,
\label{CanD-eqn-spinR}
\end{align} 
%\begin{align*}
%\partial_t\tilde\bs+{\operatorname{div}}\big(\boldsymbol{J}+\upalpha D^{-1}({\tilde\bs\times\bz})\tilde\bs\big)
%=&\ \upalpha D^{-1}\tilde\bs\times\big(\bz\times\bM
%+
%2\nabla\tilde s_z-2\bz{\operatorname{div}}\tilde\bs
%\big)
%\\
%=&\ \upalpha D^{-1}\tilde\bs_\perp\times\big(\bz\times\bM
%+
%2\nabla\tilde s_z-2\bz{\operatorname{div}}\tilde\bs_\perp
%\big)
%\end{align*}
%so that
%\begin{align*}
%\partial_t\tilde{s}_z+{\operatorname{div}}\big(\boldsymbol{J}_z+\upalpha D^{-1}({\tilde\bs_\perp\times\bz})\tilde{s}_z\big)
%=&\,
% \upalpha D^{-1}\bz\cdot\tilde\bs_\perp\times\big(\bz\times\bM
%+
%2\nabla\tilde s_z
%\big)
%\\
%=&\,
% -\upalpha D^{-1}\tilde\bs_\perp\cdot\bz\times\big(\bz\times\bM
%+
%2\nabla\tilde s_z
%\big)
%\\
%=&\,
% \upalpha D^{-1}\big(\tilde\bs_\perp\cdot\bM
%-2\tilde\bs_\perp\cdot\bz\times
%\nabla\tilde s_z
%\big)
%\\
%=&\,
% \upalpha D^{-1}\big(\tilde\bs_\perp\cdot\bM
%-2\nabla\tilde s_z\cdot\tilde\bs_\perp\times\bz
%\big)
%\\
%=&\,
%m \upalpha \tilde\bs_\perp\cdot\bv
%-2 \nabla\tilde s_z\cdot(\bu-\bv)
%\end{align*}
%\begin{align*}
%\boldsymbol{J}=&\ D\bz\big(  \bv s_z+ m^{-1}
%\bz\cdot\nabla\bs\times\bs\big)+D
%\big(  \bv \bs_\perp- m^{-1}\bz\times\bz\times
%(\nabla\bs\times\bs)\big)
%\\
%=&\ D\bz\big(  \bv s_z+ m^{-1}
%\nabla\bs_\perp\cdot\bs_\perp\times\bz\big)+D
%\big(  \bv \bs_\perp+ m^{-1}(s_z\nabla\bs_\perp-\bs_\perp\nabla s_z)\times\bz\big)
% \end{align*}
%
%
where {\color{black}\eqref{spinCurr} yields} $m(\mathbbmslbf{J}\times\bz)_j=m\bv(\tilde\bs\times\bz)_j+D^{-1}(\tilde{s}_j\nabla \tilde{s}_z- \tilde{s}_z\nabla \tilde{s}_j)$.
Once again, we can observe the effect of the anomalous velocity $\upalpha \bs\times\bz$ in the transport terms and the resulting spin Hall current $\color{black}\mathbbmslbf{J}_{\rm H}=\upalpha D^{-1}({\tilde\bs\times\bz})\tilde\bs$. The latter contributes to the new definition {\color{black}$\mathbbmslbf{J}^{tr}=\mathbbmslbf{J}+\mathbbmslbf{J}_{\rm H}$ of spin transport current in \eqref{trspicurrent}}.
{\color{black}As for the right-hand side of \eqref{Canrhoeqn-spinR}, the spin torques again include a semiclassical SOC term and three other precession} terms associated to spin-orbit correlations. Also, we observe that, unlike the  position average ${\int} xD\,\de^3x$ and the  momentum average ${\int} \bM\,\de^3x$, the average spin vector  ${\int}\tilde\bs\,\de^3x$ develops nonplanar components. This is another well known peculiar feature of SOC in Rashba dynamics. Instead, the vertical component  $\bz\cdot{\int} (\bx\times\bM+\tilde\bs)\,\de^2x$ of the total angular momentum is conserved by symmetry arguments in the absence of the external potential $V$. Notice that in the orbital momentum equation \eqref{CanMomeqn-spinR} the stress tensor $\mathbbmslbf{T}=\mathbbmslbf{T}_{\rm QGT}+\mathbbmslbf{T}_{\rm SOC}$ is given by the sum of two terms: the first term $\mathbbmslbf{T}_{\rm QGT}=-m^{-1}\tilde\bs\cdot\nabla (D^{-1}\partial_j\tilde\bs)$ is associated to quantum correlations induced by the QGT in \eqref{QGT}, while the second term $\mathbbmslbf{T}_{\rm SOC}=m\upalpha
\mathbbmslbf{J}\times\bz$ is associated to  SOC-induced quantum correlations.

{\color{black}
A simplified version of the Madelung-Rashba system can be obtained by resorting to the one-dimensional approximation of the Rashba SOC operator, that is used for quantum nanowires \cite{Mireles}. In this case, the momentum operator is $\hat{p}=-i\hbar\partial_x$ and the SOC vector field \eqref{Xdef} is replaced by
$
\bhX={\alpha_R}\widehat\sigma_z/\hbar
$,
so that $\langle\bhX\rangle=\upalpha s_z$.
In this case, $\mathbbmslbf{J}=D\big(  v \bs- m^{-1}\bs\times
\partial_x\bs\big)$ and the Madelung-Rashba equations simplify considerably. For example, in terms of the variables $D$, $\tilde\bs$, and $v=u-\upalpha \tilde{s}_z/D$ we have
\begin{align}\label{veqn-spinR1D}
&mD\Big(\frac{\partial v}{\partial t}+u\frac{\partial v}{\partial x}\Big)=
-D\frac{\partial V}{\partial x}
+ \frac1m\frac{\partial }{\partial x}\Big(\tilde\bs\cdot\frac{\partial }{\partial x}\Big(\frac{1}D\frac{\partial \tilde\bs}{\partial x}\Big)  -\frac{m\upalpha}D\bz\cdot\frac{\partial \tilde\bs}{\partial x}\times\tilde\bs\Big),
\\
&\label{seqn-spinR1D}
\frac{\partial \tilde\bs}{\partial t}+\frac{\partial }{\partial x}\Big(\mathbbmslbf{J}+\frac\upalpha{D}\tilde{s}_z\tilde\bs\Big)=
{\upalpha}\Big(mv\bz+\bz\times\frac{\tilde\bs}{D}+\bz\times\frac{\partial }{\partial x}\frac{\tilde\bs}{D}\Big)\times\tilde\bs,
\\\label{Deqn-spinR1D}
&\frac{\partial D}{\partial t}+\frac{\partial }{\partial x}(Dv+\upalpha\tilde{s}_z)=0.
\end{align}
We notice that the Hall spin current \eqref{HallSC} now reads $ \mathbbmslbf{J}_{\rm H}=\upalpha D^{-1}\tilde{s}_z\tilde\bs$, to be contrasted with the conventional expression $D\hbar\langle\{\bhX,\widehat\bsigma\}\rangle/4=\upalpha\hbar D\bz/4$, which points vertically. Also, we observe that  the quantum terms in the angular velocity $\bomega={\upalpha}(mv\bz+\bz\times{\tilde\bs}/{D}+\bz\times{\partial_x }({\tilde\bs}/{D}))$ of the spin gyration are strictly planar.
}
%\comment{write the 1D eqns?
%\[
%(\partial_t+\bu\cdot\nabla)\bv=
%-\frac1m{\nabla}(V-V_Q)
%-\frac1D\partial_j {\operatorname{Tr}}\bigg( \widehat{X}_j\MCO+\frac{\hbar^2}{2m^2}D\partial_j\hat\rho\nabla\hat\rho  \bigg).
%\]
%\[
%{\operatorname{Tr}}( \widehat{X}_j\MCO)=\frac{\alpha_R}\hbar\frac{D}{2m}{\operatorname{Tr}}( \widehat\bsigma\cdot\bz [\hat\rho,i\hbar\hat\rho'])
%=
%\frac{\alpha_R}{\hbar^2}\frac{D}{m}{\operatorname{Tr}}( \widehat\bsigma\cdot\bz  \widehat{\sigma}_a
%(\bs'\times\bs)_a)
%%=
%%\frac{2\alpha_R}{\hbar^2}\frac{D}{m}
%%(\bs'\times\bs)_z
%=\frac\upalpha{m}D(\bs'\times\bs)_z
%\]
%so that
%\[
%mD(\partial_t+u\cdot\nabla)v=
%-D(V'-V_Q')
%- \left( \upalpha D(\bs'\times\bs)_z+m^{-1}D|\bs'|^2  \right)'.
%\]
%%%
%\[
%i\hbar \partial_t\tilde\bs+{i\hbar}\Big(\mathbbmslbf{J}+\frac\upalpha{D}\tilde{s}_z\tilde\bs\Big)'=
%{\upalpha}\left(mv\bz+\bz\times\frac{\tilde\bs}{D}+\bz\times\Big(\frac{\tilde\bs}{D}\Big)'\right)\times\tilde\bs
%%\frac1D
%%\bigg[\bM\cdot\widehat{\bf X}+{i\hbar}
%%%{\color{blue}\Big[\nabla\tilde\rho+\tilde\rho\nabla {\ln}\sqrt{D},\widehat{\bX}\Big]}
%%%\qquad
%%\big[
%%\nabla\tilde\rho,\widehat{\bX}\big]-\frac{i\hbar}{2D}\big[\tilde\rho,\operatorname{div}( D\widehat{\bX})\big]
%%,\tilde\rho\bigg]
%\]
%}

\section{The \emph{bohmion}  particle closure with spin-orbit coupling\label{sec:bohmions}}
As we have seen, the hydrodynamic equations of SOC present several intricacies beyond the challenges already posed by the  terms associated to {\color{black}the} QGT. Indeed, the presence of the {\color{black}Mead spin current} leads to various terms that persist even for  planar  Rashba dynamics. 

{\color{black}In an attempt  to lay the ground for possible numerical implementations,} 
%In an attempt to make these equations tractable from a numerical viewpoint, 
here we extend the particle \emph{bohmion} scheme introduced in \cite{FoHoTr19,HoRaTr21} within the context of Madelung hydrodynamics and its extensions to molecular systems. The general idea is to formulate a regularization of  the action principle underlying SOC hydrodynamics in such a way to restore point particle solutions, which are otherwise prevented by the {\color{black}spin gradient terms in the continuum equations}. To this purpose, we perform the following replacements where appropriate: 
\beq\label{bars-vars}
D\to \bar{D}=K*D
\,,\qquad\text{and}\qquad
\tilde{\bs}\to\bar{\bs}=K*\tilde{\bs}.
\eeq
Here, $*$ denotes convolution and $K=K(\bx-\bx')$ is a translation-invariant mollifier that depends on a lenghtscale parameter $\Delta$ such that $K$ tends to a delta function in the limit $\Delta\to0$, thereby recovering the original system as a limit process. In the absence of SOC, the method was successfully benchmarked in \cite{HoRaTr21} by using a Gaussian profile for $K$ with variance $\Delta$. A variant of this method was recently proposed in \cite{BaBeGBTr24} to model the dynamics of interacting quantum and classical systems.

{\color{black}In order to retain basic consistency properties such as energy and momentum balance, we  use the regularization procedure in Hamilton's variational principle rather than in the equations of motion}.
As a first step, we write the variational principle \eqref{actprinLP1}-\eqref{actprinLP2} in the spin vector formalism as
%\beq
%\delta{\int_{t_1}^{t_2}}\ell(\bu,D,\boldsymbol\xi,\tilde\bs)\de t=0
%\quad\ \text{with}\quad\ 
%\ell= {\int} \bigg(\frac{m}{2D}\bigg|D\bu-{\tilde\bs}\times\boldsymbol{\cal F}\bigg|^2-\frac{\hbar^2}{8m}{|\nabla
%   D|^2}+\tilde\bs\cdot\boldsymbol\xi- D\mathscr{E}(\tilde\bs,\nabla\tilde\bs)\bigg)\de^3x
%\label{actprinEPs},
%\eeq
\beq\label{actprinLP1-s}
\delta{\int_{t_1}^{t_2}}\bigg({\int}\big(mD\bv\cdot\bu+\tilde{\bs}\cdot\boldsymbol\xi\big)\de^3x-h\bigg)\de t=0,
\eeq
where {\color{black}the total energy is given by replacing $\bM=mD\bv$ in \eqref{actprinLP2spin}}.
%:
%\beq
%h(\bv,D,\tilde\bs)= {\int} \bigg({mD}\frac{|\bv|^2}2+{\color{black}e\bv\cdot\tilde\bs\times\boldsymbol{\cal E}}
%%+\frac{\hbar^2}{8m}\frac{|\nabla D|^2}D
%   + {V} +{\color{black}e\frac{e}{mD}\boldsymbol{\cal E}}\cdot\big(\tilde\bs\cdot\nabla \tilde\bs- \tilde\bs\operatorname{div}\tilde\bs\big)+ \frac{\|\nabla\tilde\bs\|^2}{2mD}\bigg)\de^3x,
%\label{actprinLP-s}
%\eeq
%which is obtained from \eqref{actprinLP2}-\eqref{neweps2} by using $\tilde\rho=(D\boldsymbol{1}+2\hbar^{-1}\tilde\bs\cdot\widehat{\bsigma})/2$.
%Upon recalling $\bv=m^{-1}\bM/D$, 
Then, the equations \eqref{MomEqn-spin} and \eqref{Canrhoeqn-spin}-\eqref{CanD-eqn-spin} are obtained by taking arbitrary variations $\delta\bv$ and
\beq
\delta D=-\operatorname{div}(D\bw)
,\qquad 
\delta { \tilde\bs} = \boldsymbol\zeta\times   \tilde\bs -\operatorname{div}(\bw\tilde\bs)
,\qquad 
\delta \boldsymbol{\xi} = \partial _t \boldsymbol\zeta +  \boldsymbol\zeta \times \boldsymbol{\xi}   + \bu  \cdot\nabla  \boldsymbol\zeta   -  \bw\cdot\nabla   \boldsymbol{\xi}, 
\label{vars3-s}
\eeq
where $\bw$ and $\boldsymbol\zeta$ are arbitrary and vanish at the endpoints. As customary in Euler-Poincar\'e theory \cite{HoMaRa98,HoScSt09}, these variations are accompanied by the auxiliary transport equations
\beq\label{aux-eqns}
\partial_t D+\operatorname{div}(D\bu)=0
\,,\qquad\text{and}\qquad
\partial_t \tilde\bs+\operatorname{div}(\bu\tilde\bs)=\boldsymbol\xi\times\tilde\bs.
\eeq
At this point, we perform the replacements \eqref{bars-vars} in all the terms containing $\nabla\tilde\bs$. This amounts to replacing the total energy  \eqref{actprinLP2spin} by its regularized version
\[
\bar{h}(\bv,D,\tilde\bs)= {\int} \bigg({mD}\frac{|\bv|^2}2+{\color{black}e \bv\cdot\tilde\bs\times\boldsymbol{\cal E}}
%+\frac{\hbar^2}{8m}\frac{|\nabla D|^2}D
   + {V} +{\color{black}\frac{e}{m\bar{D}}\boldsymbol{\cal E}}\cdot\big(\bar\bs\cdot\nabla \bar\bs- \bar\bs\operatorname{div}\bar\bs\big)+ \frac{\|\nabla\bar\bs\|^2}{2m\bar{D}}\bigg)\de^3x
\]
in \eqref{actprinLP1-s}. Then, if the mollifier $K$ is regular enough, the resulting system allows for the singular solutions \cite{HoTr09}
\beq\label{singsol}
D(\bx,t)=\sum_{a=1}^Nw_a\delta(\bx-\bq_a(t))
\,,\qquad\qquad
\tilde{\bs}(\bx,t)=\sum_{a=1}^Nw_a\bmu_a(t)\delta(\bx-\bq_a(t)),
\eeq
where $w_a>0$ and $\sum_a w_a=1$. Also, enforcing $\sum_a |\bmu_a|^2=\hbar^2/4$ ensures that $|\int\tilde\bs\,\de^3x|\leq\hbar/2$ via $|\sum_a w_a \bmu_a|^2\leq (\sum_a w_a^2)(\sum_a |\bmu_a|^2)\leq (\sum_a w_a)^2(\sum_a |\bmu_a|^2)$, where the last inequality follows from $w_a>0$.

%Notice that inserting the first of \eqref{singsol} in \eqref{D-eqn-spin} yields $\dot\bq_a(t)=\bu(\bq_a(t),t)$.
{\color{black}At this point, we insert the ansatzes \eqref{singsol}  in} the variational principle $\delta{\int_{t_1}^{t_2}}\big({\int}(mD\bv\cdot\bu+\tilde{\bs}\cdot\boldsymbol\xi)\,\de^3x-\bar{h}\big)\de t=0$, which then becomes, upon denoting $\bp_a(t)=m\bv(\bq_a(t),t)$,
\begin{equation}\label{bohmion-varprin}
\delta{\int_{t_1}^{t_2}}\bigg(\sum_{a=1}^Nw_a\big(\bp_a\cdot\dot{\bq}_a+\bmu_a\cdot\boldsymbol\xi_a\big) -\mathsf{h}\big(\{\bq_a\},\{\bp_a\},\{\bmu_a\}\big)\bigg)\de t=0,
\end{equation}
where
\begin{multline}\label{bohmionHam}
\mathsf{h}\big(\{\bq_a\},\{\bp_a\},\{\bmu_a\}\big)=
\frac{|\bp_a|^2}{2m}+{\color{black}\frac{e}m\bp_a\cdot\bmu_a\times\boldsymbol{\cal E}_{\!a\,}}
   +{V}_a
\\ +\frac1{2\color{black}m}\sum_{b=1}^Nw_b\bigg({\color{black}e}{\bmu_a\times\bmu_b}\cdot{\int}\frac{\boldsymbol{\cal E}\times( K_a\nabla  K_b-K_b\nabla K_a \big)}{\sum_c w_c K_c}\,\de^3x+ {\bmu_a\cdot\bmu_b}{\int}\frac{\nabla K_a\cdot\nabla K_b}{\sum_c w_c K_c}\,\de^3x\bigg).
\end{multline}
Here, we have used the fact that inserting the first of \eqref{singsol} in the first of \eqref{aux-eqns} yields
$\dot\bq_a(t)=\bu(\bq_a(t),t)$. Also, we have introduced the notation  $\boldsymbol\xi_a=\boldsymbol\xi(\bq_a)$, $\color{black}\boldsymbol{\cal E}_{\!a\,}=\boldsymbol{\cal E}(\bq_a)$, $V_a=V(\bq_a)$, and $K_a=K(\bx-\bq_a)$. In \eqref{bohmion-varprin}, the variations $\delta\bp_a$ and $\delta\bmu_a$ are arbitrary, while the  relation $\delta\boldsymbol\xi_a(t)=\delta \boldsymbol\xi(\bq_a(t),t)$ leads to
\[
\delta \boldsymbol\xi_a =  \dot{ \boldsymbol\zeta}_a +  \boldsymbol\zeta_a \times \boldsymbol{\xi}_a,
\] 
{\color{black}with arbitrary $\boldsymbol\zeta_a$}.
At this point, we obtain a finite-dimensional Hamiltonian system in the variables $(\{\bq_a\},\{\bp_a\},\allowbreak\{\bmu_a\})$, which  obey the following equations:
\beq\label{bohmion-eqns}
\dot{\bq}_a=\frac1{w_a}\frac{\partial \sf h}{\partial \bp_a}
\,,\qquad\qquad
\dot{\bp}_a=-\frac1{w_a}\frac{\partial \sf h}{\partial \bq_a}
\,,\qquad\qquad
\dot{\bmu}_a=\frac1{w_a}\frac{\partial \sf h}{\partial \bmu_a}\times\bmu_a.
\eeq
Due to their construction, these equations inherit a Hamiltonian structure comprising the Hamiltonian \eqref{bohmionHam} and the direct-sum Poisson bracket
\begin{align*}
	\{\!\!\{h,k\}\!\!\}\big(\{\bq_a\},\{\bp_a\},\{\bmu_a\}\big)
	=\sum_{a=1}^N\frac1{w_a}\left({\color{black}\frac{\partial h}{\partial \bq_a}\cdot\frac{\partial k}{\partial \bp_a}-\frac{\partial h}{\partial \bp_a}\cdot\frac{\partial k}{\partial \bq_a}}+\bmu_a\cdot\frac{\partial h}{\partial\bmu_a}\times\frac{\partial k}{\partial\bmu_a}
\right).	
\end{align*}
Therefore, equations \eqref{bohmion-eqns} retain exact balance laws for energy and momentum from the original continuum model. 
Their implementation {\color{black}would require computing different} integral terms at each time step. For example, the equation of the spin variables reads
\begin{multline}\label{mueqn}
\dot{\bmu}_a=-{\color{black}\frac{e}m\bmu_a\times(\boldsymbol{\cal E}_{\!a\,}\times\bp_a)}
\\
-\frac{\bmu_a}m\times\sum_{b=1}^Nw_b\bigg(e{\bmu_b}\times{\int}\frac{\boldsymbol{\cal E}\times( K_a\nabla  K_b-K_b\nabla K_a \big)}{\sum_c w_c K_c}\,\de^3x+{\bmu_b}{\int}\frac{\nabla K_a\cdot\nabla K_b}{\sum_c w_c K_c}\,\de^3x\bigg).
\end{multline}
%\comment{Do I want to say something in 1D?
%\begin{multline}%\label{bohmionHam}
%\mathsf{h}\big(\{q_a\},\{p_a\},\{\bmu_a\}\big)=
%\frac{p_a^2}{2m}+\upalpha p_a\bmu_a\cdot\bz
%   +{V}_a
%\\ +\frac1{2\color{black}m}\sum_{b=1}^Nw_b\bigg(\upalpha\bz\cdot{\bmu_a\times\bmu_b}{\int}\frac{( K_a  K_b'-K_b K_a' \big)}{\sum_c w_c K_c}\,\de^3x+ {\bmu_a\cdot\bmu_b}{\int}\frac{ K_a'\cdot K_b'}{\sum_c w_c K_c}\,\de^3x\bigg).
%\end{multline}
%}\noindent
Instead, the velocity of each trajectory reads $\dot{\bq}_a=\color{black}(\bp_a+e\bmu_a\times\boldsymbol{\cal E}_{\!a\,})/m$, whose second term represents the anomalous velocity contribution.  {\color{black}Notice that, as the variance of the Gaussian kernel $K$ tends to infinity, its derivatives tend to zero and the present closure recovers the mean-field equations.

The forces and torques originating from the first  integral in \eqref{bohmionHam} represent an extension of the bohmion scheme previously implemented and benchmarked in \cite{HoRaTr21}, which only retains the second energy integral associated to the QGT. Instead, the first energy integral generates SOC-induced correlations whose numerical effects remain to be tested  in the future.
Here, we notice that, as already observed in \cite{HoRaTr21}, each trajectory is simultaneously coupled to all others via both integral energy terms in \eqref{bohmionHam}, thereby reflecting the nonlocal interactions generally associated to quantum correlations. Each of the energy integrals leads to a different behavior and carries a different level of complexity: while the SOC terms carry only one derivative, the QGT terms carry two derivatives.}
%This is a feature of quantum correlations that manifest in this scheme via the integrals} representing nonlocal interactions. We see that, once again, different types of quantum correlations are associated to both the SOC terms carrying the force field $\color{black}e\boldsymbol{\cal E}/m$ and  the QGT. We also observe that each of these terms leads to a different behavior and carries a different level of complexity: while the SOC terms carry only one derivative, the QGT terms carry two derivatives. 
This SOC-extension of the bohmion scheme in \cite{FoHoTr19,HoRaTr21} may be helpful in probing the different mechanisms associated to SOC and QGT correlations by switching on and off the corresponding terms. For example, this can be done in first place by considering planar Rashba dynamics for which $\color{black}\boldsymbol{\cal E}=m\upalpha\bz/e$. {\color{black}As a further simplification, one may also consider the one-dimensional approximation as in equations \eqref{veqn-spinR1D}-\eqref{Deqn-spinR1D}. In this case, the  Hamiltonian \eqref{bohmionHam} becomes 
\begin{multline*}%\label{bohmionHam}
\mathsf{h}\big(\{q_a\},\{p_a\},\{\bmu_a\}\big)=
\frac{p_a^2}{2m}+\upalpha p_a\bz\cdot\bmu_a
   +{V}_a
\\ +\frac1{2m}\sum_{b=1}^Nw_b\bigg(\upalpha\bz\cdot{\bmu_a\times\bmu_b}{\int}\frac{ K_a'  K_b-K_b' K_a}{\sum_c w_c K_c}\,\de x+ {\bmu_a\cdot\bmu_b}{\int}\frac{ K_a' K_b'}{\sum_c w_c K_c}\,\de x\bigg),
\end{multline*}
and, for example, the equation \eqref{mueqn} for the spin variables is replaced by
\[
\dot{\bmu}_a=\upalpha p_a\bz\times\bmu_a
-{\bmu_a}\times\sum_{b=1}^Nw_b\bigg(\upalpha{\bmu_b}\times\bz{\int}\frac{ K_a'  K_b-K_b' K_a}{\sum_c w_c K_c}\,\de x+\frac{\bmu_b}m{\int}\frac{ K_a' K_b'}{\sum_c w_c K_c}\,\de x\bigg).
\]
As in the higher-dimensional cases, we observe the appearance of both the SOC-induced and the QGT-induced integrals identifying the different types of correlation effects. While the latter was successfully implemented in \cite{HoRaTr21}, the implementation of the SOC-induced integral and its associated forces and torques}
is left for future work.

%\section{Analogies with mixed quantum-classical models}
%
%This section explores certain analogies between quantum SOC dynamics and recent models of for the interaction dynamics of quantum and classical systems. These analogies are made available by the appearance of the MCO in both contexts.
%
%\dots\vspace{3cm}

\section{Conclusions and outlook}

While SOC has been  known for about a century \cite{SpMa15}, the various aspects of its intricate phenomenology still lack a unified framework capable of elucidating their close dynamical interplay. This paper establishes such a framework by leveraging the action principles underlying the corresponding  hydrodynamic formulation, naturally bringing together spin and orbital quantities. 
By isolating the role of SOC in the absence of magnetic fields and further relativistic corrections, our treatment {\color{black}reveals a decomposition of the  spin current 
  \eqref{spincurrent} into semiclassical and  correlation terms. This decomposition} proves essential for characterizing the different types of SOC-induced orbital forces and spin torques, effectively  distinguishing  them from the QGT contributions  governing quantum correlations in the absence of SOC.

This level of investigation is made possible by the use of the variational principles and the Hamiltonian structure underlying Madelung hydrodynamics with spin. Well established in geometric mechanics \cite{HoScSt09,MaRa98}, these mathematical structures provide a formal distinction between kinematic effects and dynamical forces, thereby leading to a physical characterization of the various terms in the final equations of motion. For example, this approach directly unfolds  the role of the anomalous velocity in the  current shift associated to the spin Hall effect. Indeed,  the anomalous velocity contribution to the spin transport dynamics leads to the new expression \eqref{trspicurrent} of the overall transport spin current, which emerges naturally in Madelung hydrodynamics thereby addressing an open question  in spintronics \cite{Ra03,TaKaMi24}. 
Likewise, the anomalous velocity was shown to have a crucial role also in the orbital dynamics and the evolution of the spin geometric phase. {\color{black}At the experimental level, the actual measurement of transport spin current effects in a pure Rashba medium is deemed questionable \cite{Samokhin,Sonin3}. However, when this system is coupled to a ferromagnet under a longitudinal electric field, the transverse anomalous velocity shift drives a robust spin accumulation at the spatial boundary, thereby exerting a measurable precessional spin-orbit torque \cite{Kim} that rotates a macroscopic magnetization. In this context, evaluating the  macroscopic predictions of this new approach may offer a concrete path to benchmark our variational framework against previous spin transport formulations in  experimental configurations. The preliminary investigation in the Appendix shows that the Fermi-Dirac equilibrium average of the new transport spin current   scales linearly with the Fermi energy {\color{black}or, in other words, with the electron concentration \cite{Samokhin}. Instead, the latter   is entirely} absent in the case of the conventional definition.

While the density matrix representation provides a  transparent treatment that can be adapted to different contexts and naturally distinguishes between spin algebra operations and differential operators in physical space, the spin vector formalism offers a direct quantitative approach to spin transport features. The second part of this paper follows the latter approach to present the new definition of the transport spin current and formulate the Madelung-Rashba system for planar hydrodynamics with Rashba SOC. In this case, the orbital momentum equation simplifies substantially while the spin torques reflect the complexity of quantum correlations. {\color{black}A further simplification is  made available by the one-dimensional approximation of the Rashba operator as used in quantum nanowires. In this case, the complexity level is substantially lower}.

The paper concludes by proposing  a particle {\color{black}closure for possible numerical implementations. Based on the variational principle of the original model, this {\color{black}closure} allows a natural separation between SOC and QGT correlation terms in such a way that they can be studied separately by switching on and off the corresponding terms in the Hamiltonian function. This study of the effects arising from different correlation terms is a most interesting direction that will shed new light on the {\color{black}mechanisms of quantum correlation   dynamics}.

We are also interested in further developing the link between the MCO and analogous quantities appearing prominently in quantum-classical interaction dynamics. Indeed, the SOC term occurring in \eqref{neweps} has a direct analogue in the variational principles governing recent models of quantum-classical coupling  \cite{BaBeGBTr24,GBTr21,GBTr22,GBTr24}.  {\color{black}These models overcome the consistency issues of semiclassical Boltzmann equations commonly used in spintronics \cite{Kailasvuori,Halperin,Raimondi}. For example,  the latter fail to retain positivity of both classical probability and the quantum density matrix, thereby  invalidating the Heisenberg principle. This point was recently raised in \cite{BeMaTr26}, where the new consistent quantum-classical models were used to study Rashba nanowires. 
 The occurrence of direct analogies between SOC and quantum-classical coupling} and the predominant role of the MCO in both settings suggests a unified geometric origin across disparate theoretical frameworks and can shed light on the nature of quantum-classical correlations in  hybrid systems.

\paragraph{Acknowledgments.} The author is grateful to  Fran\c{c}ois Gay-Balmaz, {\color{black}Eran Ginossar, Cosimo Gorini}, Darryl Holm, Giovanni Manfredi, {\color{black}Rocco Martinazzo}, Branislav Nikoli\'c, {\color{black}Edouard Sonin, and Giovanni Vignale} for several discussions and correspondence on this and related topics during the course of this work. Financial support by the Leverhulme Trust Research Project Grant RPG-2023-078 is greatly acknowledged.

{\color{black}
\appendix

\section{\makebox{Equilibrium  spin current in the Rashba medium}}

This appendix follows the procedure in \cite{Sonin1} to present equilibrium relations for the transport spin current in \eqref{trspicurrent}.  Let $\bz$ denote the unit vertical and let us consider  the Rashba Hamiltonian
\[
\widehat{H}=\frac{\hbar^2}{2m}\hat{p}^2+\frac{\alpha_R}\hbar\widehat{\bsigma}\times\bz\cdot\hat{\bp},
\]
with eigenstates 
\[
\Psi_\pm=\frac1{\sqrt{2}}\left(\begin{matrix}1 \\ \pm \frac{k_y-ik_x}{k}\end{matrix}\right)e^{i\bk\cdot\bx},
\]
so that so that $D=\Psi^\dagger_\pm\Psi_\pm=1$ and the spin density reads
\[
\langle\widehat{\bs}\rangle_\pm=\frac\hbar2\Psi_\pm^\dagger\widehat{\bsigma}\Psi_\pm
=\mp\frac\hbar{2k}\bz\times{\bk}.
\]

Here, we compute the  zero-temperature Fermi-Dirac averages of the transport spin current. For the two-band and single-band averages, we  use the formulas \cite{Sonin1}
\[
\langle \mathbbmslbf{J}^{tr}\rangle = \frac1{(2\pi)^2}\sum_\pm{\int_{k<k_{F\pm}}}\mathbbmslbf{J}^{tr}_\pm\,\de^2\bk
= \frac1{(2\pi)^2}\sum_\pm{\int_0^{2\pi}}\de\theta{\int_0^{k_{F\pm}}}
k\de k\,\mathbbmslbf{J}^{tr}_\pm
\]
and
\[
\langle \mathbbmslbf{J}^{tr}\rangle = \frac1{(2\pi)^2}{\int_{-k_{F+}<k<k_{F-}}}\mathbbmslbf{J}^{tr}_-\,\de^2\bk
= \frac1{(2\pi)^2}{\int_0^{2\pi}}\de\theta{\int_{-k_{F+}}^{k_{F-}}}
k\de k\,\mathbbmslbf{J}^{tr}_-,
\]
respectively. Here, $\mathbbmslbf{J}^{tr}_\pm$ is the transport spin current evaluated on the $\pm$-subband, while $k_{F\pm}=k_m\mp m\alpha_R/\hbar^2$ is known as the \emph{subband Fermi wavevector} and $k_m$ as the \emph{average Fermi wavevector}. Also, $\varepsilon_F=\hbar^2(k_m^2-m^2\alpha_R^2/\hbar^4)/2$ is the Fermi energy, that is the chemical potential at zero-temperature.
We  evaluate the  definition \eqref{trspicurrent} of the tranport spin current on the eigenstates $\Psi_\pm$, that is
\begin{align*}
\mathbbmslbf{J}^{tr}_\pm=&\,
\frac{\hbar}{m}\bk\langle\widehat{\bs}\rangle_\pm-\frac{2\alpha_R}{\hbar^2}(\langle\widehat{\bs}\rangle_\pm\times\bz)\langle\widehat{\bs}\rangle_\pm
%\\
%&\,
%=\mp\frac{\hbar^2}{2mk}\bk(\bz\times{\bk})
%+
%\frac{\alpha_R}{2 k^2}(\bz\times\bz\times\bk)(\bz\times{\bk})
%\\
%&\,
%=\mp\frac{\hbar^2}{2mk}\bk(\bz\times{\bk})
%-
%\frac{\alpha_R}{2 k^2}\bk(\bz\times{\bk})
%\\
%&\,
%=
%\frac{\hbar^2}{2mk}\left(\mp1
%-\frac{\alpha_S}{ k}\right)\bk
%\left(\bz\times{\bk}\right)
\\
&\,
=
-\frac{\hbar^2}{2mk}\left(\pm1
+\frac{\alpha_S}{ k}\right)\bk
(\bz\times{\bk}),
\end{align*}
where we have denoted $\alpha_S=m\alpha_R/\hbar^2$. Equivalently,  in polar coordinates, 
%\[
%\bk
%\left(\bz\times{\bk}\right)=k^2\left(\begin{matrix} 
%- \sin\theta\cos\theta &  \cos^2\theta  \\ 
%- \sin^2\theta & \sin\theta\cos\theta 
%\end{matrix}\right)
%=
%-k^2\sin\theta\cos\theta\widehat{\sigma}_z+k^2\left(\begin{matrix} 
%0 &  \cos^2\theta  \\ 
%- \sin^2\theta & 0
%\end{matrix}\right)
%\]
%and thus
\[
\mathbbmslbf{J}^{tr}_\pm=
-\frac{\hbar^2}{2m}\left(\pm k
+{\alpha_S}\right)
\left(
-\sin\theta\cos\theta\widehat{\sigma}_z+\left(\begin{matrix} 
0 &  \cos^2\theta  \\ 
- \sin^2\theta & 0
\end{matrix}\right)
\right).
\]

\paragraph{Two-band average.}
Introducing the matrix $\mathbbm{K}_{jh}=\upvarepsilon_{jh\ell}  z_\ell$, the two-band Fermi average reads
\begin{align*}
\langle \mathbbmslbf{J}^{tr}\rangle 
= &\,
-
\frac{\hbar^2}{2m(2\pi)^2}{\int_0^{2\pi}}\de\theta\left(\begin{matrix} 
0 &  \cos^2\theta  \\ 
- \sin^2\theta & 0
\end{matrix}\right)\sum_\pm
{\int_0^{k_{F\pm}}}
k\de k\,\left(\pm
k+{\alpha_S}\right)
\\= &\,
-
\frac{\pi\hbar^2}{2m(2\pi)^2}\left(
{\int_0^{k_m+\alpha_S}}
k\de k\,\left(-
k+{\alpha_S}\right)
+
{\int_0^{k_m-\alpha_S}}
k\de k\,\left(
k+{\alpha_S}\right)
\right)\mathcal{E}
%\\=&\,
%-\frac{\pi\hbar^2}{6m(2\pi)^2}\alpha_S\left( \alpha_S^2-3k_m^2\right)\mathbbm{K}
\\=&\,
\frac{\hbar^2\alpha_S}{4m\pi}\left(k_m^2- \frac{\alpha_S^2}3\right)\mathbbm{K}
\\=&\,
\frac{\alpha_S}{4\pi m}\left( {\varepsilon_F}+\frac{\hbar^2}3\alpha_S^2\right)\mathbbm{K}
%\\=&\,
%\frac{m\alpha_R}{4\pi \hbar^2}\left( \frac{\varepsilon_F}{m}+\frac{m}{3\hbar^2}\alpha_R^2\right)\mathbbm{K},
\end{align*}
where we recall $
\varepsilon_F={\hbar^2}(k^2_m-\alpha_S^2)/2
$. This result needs to be contrasted with the corresponding result associated to the conventional definition \eqref{convSC}, that is \cite{Sonin1}
\[
\langle \mathscr{J}\rangle= - \frac{\hbar^2\alpha_S^3}{6m\pi}\mathbbm{K}.
\]
While the conventional definition yields a  result that  ignores the energy level of the system, the definition \eqref{trspicurrent} indicates that the baseline spin flow scales linearly with the  Fermi energy of the Rashba medium. This dependence of the equilibrium current average on the electron concentration  is surprisingly missing in the conventional expression of $\langle \mathscr{J}\rangle$. It is interesting to compare the above expression of $\langle \mathbbmslbf{J}^{tr}\rangle $ with the analogous result recently found in \cite{Samokhin}, based on a different  definition of transport spin current. In that case (see equation (52) therein), the equilibrium current average is proportional to $\alpha_S\varepsilon_F^2$, so that in that case the linear scaling with the Rashba constant is accompanied by a quadratic scaling with the Fermi energy.

\paragraph{Single-band average.}
For the single-band regime we have
\begin{align*}
\langle \mathbbmslbf{J}^{tr}\rangle 
= &\,
\frac1{(2\pi)^2}{\int_0^{2\pi}}\de\theta{\int_{-k_F+}^{k_{F-}}}
k\de k\,\mathbbmslbf{J}^{tr}_-
\\
= &\,-\frac{\hbar^2}{2m(2\pi)^2}{\int_0^{2\pi}}\de\theta\left(\begin{matrix} 
0 &  \cos^2\theta  \\ 
- \sin^2\theta & 0
\end{matrix}\right)
{\int_{-k_m+\alpha_S}^{k_m+\alpha_S}}
k\de k\,\left(
-k+{\alpha_S}\right)
\\
= &\,-\frac{\pi\hbar^2}{2m(2\pi)^2}
{\int_{-k_m+\alpha_S}^{k_m+\alpha_S}}
k\de k\,\left(
-k+{\alpha_S}\right)\mathbbm{K}
%\\
%= &\,
%\left(-\frac{\pi\hbar^2}{2m(2\pi)^2}\right)
%\left(-\frac{2k_m^2+6\alpha_S^2}3\right)k_m\mathbbm{K}
\\
= &\,\frac{\hbar^2k_m}{4m\pi}
\left(\alpha_S^2+\frac{k_m^2}3
\right)\mathbbm{K}
%\\
%= &\,\frac{k_m}{4\pi}
%\left(\frac{\hbar^2}{3m}{k_m^2}+\frac{m}{\hbar^2}\alpha_R^2
%\right)\mathbbm{K},
.
\end{align*}
Once again, we contrast this result with the corresponding result associated to the conventional definition \eqref{convSC}, that is \cite{Sonin1}
\[
\langle \mathscr{J}\rangle= \frac{\hbar^2k_m}{4m\pi}\left(\alpha_S^2-\frac{k_m^2}3\right)\mathbbm{K}.
\]
While the average Fermi wavevector, and therefore the Fermi energy, appears in both formulas, the new definition \eqref{trspicurrent}  leads to a  positive quantity, unlike the conventional definition \eqref{convSC}.

}

\end{document}